\documentclass[fleqn,usenatbib]{mnras}

\usepackage{newtxtext,newtxmath}
\usepackage[T1]{fontenc}

\DeclareRobustCommand{\VAN}[3]{#2}
\let\VANthebibliography\thebibliography
\def\thebibliography{\DeclareRobustCommand{\VAN}[3]{##3}\VANthebibliography}

\usepackage{graphicx}	% Including figure files
\usepackage{amsmath}	% Advanced maths commands
\newcommand{\fable}{\textsc{fable}}
\title[The growth and impact of gargantuan black holes]{The growth of the gargantuan black holes powering high-redshift quasars and their impact on the formation of early galaxies and protoclusters}

\author[Bennett et al.]{
Jake S. Bennett,$^{1, 2, 3}$\thanks{E-mail: jake.bennett@cfa.harvard.edu}
Debora Sijacki,$^{2, 3}$
Tiago Costa,$^{4}$
Nicolas Laporte,$^{3,5}$
and Callum Witten$^{2,3}$
\\
$^{1}$Center for Astrophysics, Harvard University, Cambridge, MA 02138, USA\\
$^{2}$Institute of Astronomy, University of Cambridge, Madingley Road, Cambridge, CB3 0HA, UK\\
$^{3}$Kavli Institute for Cosmology Cambridge, University of Cambridge, Madingley Road, Cambridge, CB3 0HA, UK\\
$^{4}$Max-Planck-Institut f\"ur Astrophysik, Karl-Schwarzschild-Straße 1, D-85748 Garching b. M\"unchen, Germany\\
$^{5}$Cavendish Laboratory, University of Cambridge, Madingley Road, Cambridge, CB3 9BB, UK
}

\date{MNRAS, submitted}

\pubyear{2023}

\begin{document}
\label{firstpage}
\pagerange{\pageref{firstpage}--\pageref{lastpage}}
\maketitle

\begin{abstract}
High-redshift quasars ($z\gtrsim6$), powered by black holes (BHs) with large inferred masses, imply rapid BH growth in the early Universe. The most extreme examples have inferred masses of $\sim \! 10^9$\,M$_\odot$ at $z = 7.5$ and $\sim \! 10^{10}$\,M$_\odot$ at $z = 6.3$. Such dramatic growth via gas accretion likely leads to significant energy input into the quasar host galaxy and its surroundings, however few theoretical predictions of the impact of such objects currently exist. We present zoom-in simulations of a massive high-redshift protocluster, with our fiducial \textsc{fable} model incapable of reproducing the brightest quasars. With modifications to this model to promote early BH growth, such as earlier seeding and mildly super-Eddington accretion, such `gargantuan' BHs can be formed. With this new model, simulated host dust masses and star formation rates are in good agreement with existing \textit{JWST} and ALMA data from ultraluminous quasars. We find the quasar is often obscured as it grows, and that strong, ejective feedback is required to have a high probability of detecting the quasar in the rest-frame UV. Fast and energetic quasar-driven winds expel metal-enriched gas, leading to significant metal pollution of the circumgalactic medium (CGM) out to twice the virial radius. As central gas densities and pressures are reduced, we find weaker signals from the CGM in mock X-ray and Sunyaev-Zeldovich maps, whose detection - with proposed instruments such as \textit{Lynx}, and even potentially presently with ALMA - can constrain quasar feedback.
\end{abstract}


%%%%%%%%%%%%%%%%%%%%%%%%%%%%%%%%%%%%%%%%%%%%%%%%%%

%%%%%%%%%%%%%%%%% BODY OF PAPER %%%%%%%%%%%%%%%%%%

\section{Introduction}

There is now a wealth of observational evidence that links the masses of supermassive black holes with the properties of their host galaxies, such as the mass and velocity dispersion of the host's bulge \citep[see e.g.][for a review]{KormendyHo2013}. For black holes to grow to their current large masses, they must have undergone significant episodes of gas accretion in their past, in accord with Soltan's argument \citep{Soltan1982}, as all currently known mechanisms can only produce relatively small black hole seeds, with masses $\lesssim 10^6\, \mathrm{M}_\odot$ \citep{Rees1984,Volonteri2010}. It is further believed that the accretion of gas to fuel black hole growth also powers quasars, which can release huge amounts of energy into the surrounding medium \citep[for seminal papers, see][]{Schmidt1963,Salpeter1964,LyndenBell1969}. 

In the past two decades, hundreds of quasars have been detected at high redshift, $z \gtrsim 6$, suggesting that the black holes that power them must have grown to high masses, $10^{8-10}\,\mathrm{M}_\odot$, in the first billion years of the Universe's existence \citep[for a recent review see][]{Inayoshi2020}. Black holes in excess of $10^9\,\mathrm{M}_\odot$ have even been detected above $z\sim7$, further challenging theoretical models of the growth of such objects \citep[e.g.][]{Mortlock2011,Banados2018,Yang2020,Wang2021,Farina2022}. The current record holder at $z>6$ in terms of luminosity and black hole mass is SDSS J010013.02+280225.8 (henceforth referred to as J0100+2802), with a hefty inferred mass exceeding $10^{10}\,\mathrm{M}_\odot$ at $z=6.3$ \citep{Wu2015}.

The detected high-redshift quasars are typically very bright, with bolometric luminosities of $10^{47-48}$\,erg\,s$^{-1}$, implying significant energy input into their surroundings. This feedback is generally expected to explain the black hole-galaxy correlations we observe at lower redshift \citep[e.g.][]{Haehnelt1998,SilkRees1998}, however the mechanism by which energy from active galactic nuclei (AGN) couples to galaxies and clusters and their circumgalactic or intracluster media (CGM and ICM, respectively) is still not fully understood \citep[for reviews see e.g.][]{Fabian2012, KingPounds2015}. 

Powerful outflows driven by AGN have been detected around high-redshift quasars, with speeds ranging from a few hundred to more than $1000$\,km\,s$^{-1}$ and significant masses of gas being ejected from the central galaxy into the CGM \citep[e.g.][]{Maiolino2012,Bischetti2019,Marshall2023,YangJ2023}. Such outflows could not only eject gas and dust from the host galaxy and enrich the CGM around such quasars with metals, but could also heat and clear out the gaseous halo itself up, reducing the amount of inflowing gas that can reach the central galaxy to fuel star formation and further black hole accretion.

A further complication to the story of black hole growth at high redshift comes from the measurement of quasar proximity zones, which can be used to estimate the lifetime that a quasar has been actively emitting ionising radiation. This can sometimes lead to very short estimated lifetimes of just $\sim\!10^4$\,yr \citep[e.g.][]{Eilers2018,Eilers2021}, which is difficult to reconcile with large inferred black holes masses unless accretion is primarily obscured \citep[e.g.][]{Hopkins2005,Satyavolu2023,Soltinsky2023}. This is particularly true for the proximity zone around J0100+2802, whose small size implies a quasar age of $\lesssim10^5$\,yr - not long enough to grow to its observed size unless growth is hidden by dust \citep{Eilers2017,Davies2020}. The young ``age'' of J0100+2802 could also explain why no Ly$\alpha$ nebula is detected around it, as it has only recently become unobscured \citep{Farina2019}. The current and future use of \textit{JWST} will allow a much more detailed census of obscured AGN, which will help determine how much early black hole growth occurs in obscured systems \citep[e.g.][]{YangG2023}.

There have been a number of recent large-scale cosmological simulations that aim to capture such processes, and evolve populations of black holes to approximately match local constraints, such as Illustris \citep{Illustris1,Illustris2,Illustris3,Illustris4}, IllustrisTNG \citep{TNG1,TNG2,TNG3,TNG4,TNG5}, \textsc{eagle} \citep{EAGLE1,EAGLE2}, Horizon-AGN \citep{Dubois2014}, \textsc{simba} \citep{Dave2019}, and ASTRID \citep{Ni2022b} - for a comparison of black hole populations between many of these models, see \citet{Habouzit2022b}. However, the likely hosts of such bright quasars at high redshift have halo masses in excess of $\sim\!10^{12}\,\mathrm{M}_\odot$ \citep[e.g.][]{Porciani2004,Costa2023}, making them rare overdensity peaks in the box sizes of these simulations. For example, in the \textit{MassiveBlack} simulation, which has a large simulation volume of ($740$\,Mpc)$^3$, only ten black holes of $\sim\!10^9\,\mathrm{M}_\odot$ were found at $z=6$ \citep{DiMatteo2012}.  

To circumvent this numerical limitation, the zoom-in simulation technique is regularly used, which studies the evolution of these particularly rare objects by resimulating the biggest haloes from large, dark-matter-only simulations \citep[e.g.][]{Sijacki2009,Costa2014,Feng2014,Smidt2018,Lupi2019,Huang2020,Zhu2022}. Some recent works have also used a technique involving constrained Gaussian realisations to create rare overdensities within small volume simulations \citep[see e.g.][]{Ni2022a,Bhowmick2022}. Such zoom-in simulations have produced a number of supermassive black holes of $10^{9}\,\mathrm{M}_\odot$, which has also been done in sufficiently large cosmological boxes \citep{DiMatteo2012}. However, they still struggle to form some of the most massive observed black holes ($\geq10^{10}\,\mathrm{M}_\odot$) at $z \gtrsim 6$. Only a handful of these `gargantuan' black holes have been generated in previous works. In \citet{Costa2014}, a black hole with mass $\sim\!2\times10^{10}$\,M$_\odot$ was grown in a run evoking, counter-intuitively, strong galactic winds. This was due to a particularly extreme case of stellar feedback where a suppression of star formation leads to increased gas fractions in the gaseous halo, that in turn leads to enhanced black hole feeding by $z=6$. \citet{Valentini2021} grew a black hole with a mass in excess of $10^{10}\,\mathrm{M}_\odot$ by $z=7$, but only when they turned AGN feedback off. And finally \citet{Zhu2022} produced a black hole with a mass just below $10^{10}\,\mathrm{M}_\odot$ in a cosmological simulation with feedback, though this required using both a very heavy seed mass of $10^{6}\,\mathrm{M}_\odot$ and a very massive halo with a mass $>10^{13}\,\mathrm{M}_\odot$ by $z=6$. 

Difficulties in growing ultramassive black holes persist even when invoking extremely efficient gas accretion onto very massive seeds ($\sim\!10^{5-6}\,\mathrm{M}_\odot$). Under the assumption that their estimated masses are not biased high, this indicates that either these `gargantuan' black holes are rarer than what has currently been explored with zoom-in simulations, or that we are missing some fundamental knowledge about their formation and assembly. In either case, theoretical predictions of the impact of these ultramassive black holes in the early Universe is scarce, despite the growing observational evidence of their existence. Therefore the primary aim of this paper is to explore the imprints of $\gtrsim 10^{10} \,\mathrm{M}_\odot$ black holes and their progenitors on their surroundings, filling in this important gap in our theoretical understanding.  

The number of known high-redshift quasars will likely increase significantly in the coming years, due to ongoing and planned deep, wide-field surveys, such as using LOFAR in the radio \citep[e.g.][]{SmithD2016,Gloudemans2022}, \textit{eROSITA} in X-rays \citep{eROSITA}, the Vera Rubin Observatory's Legacy Survey of Space and Time (LSST) at optical wavelengths \citep{LSST}, as well as \textit{Euclid} \citep{Euclid} and the \textit{Nancy Grace Roman Space Observatory} \citep{ROMAN} in infrared. Such observations will also push these discoveries to lower luminosities too, giving a more complete picture of the build up of the black hole population in the early Universe.

The host galaxies and gaseous haloes of high-redshift quasars, as well as the quasar properties themselves, can also be investigated across the electromagnetic spectrum (and beyond). Using ALMA and \textit{JWST}, it is possible to infer properties of the host galaxies such as dust masses \citep[e.g.][]{Tripodi2023} or stellar light \citep[e.g.][]{Ding2023}. \textit{JWST} has already been used to find (candidate) accreting black holes at lower masses and even higher redshifts \citep{Furtak2023,Kocevski2023,Labbe2023,Larson2023,Maiolino2023a,Maiolino2023b,Matthee2023}. As discussed in \citet{Inayoshi2020}, it could even be possible to detect $10^5\,\mathrm{M}_\odot$ black holes accreting at the Eddington rate at $z=10$ with long exposures from NIRCam. \textit{JWST} can also be used to characterise galaxies hosting AGN detected using different methods \citep[such as in the X-ray][]{Bogdan2023,Goulding2023}. X-ray spectra from the upcoming \textit{XRISM} \citep{XRISM} and \textit{ATHENA} \citep{ATHENA}, and proposed \textit{Line Emission Mapper} \citep{LEM} will allow a detailed high-energy study of the quasar light itself, and of the quasar surroundings via absorption. And with imaging from \textit{ATHENA}, or the proposed \textit{AXIS} \citep{AXIS} and \textit{Lynx} \citep{Lynx} telescopes, it may be possible to investigate the host gaseous haloes of high-redshift quasars directly in emission. Future constraints on the power of black hole feedback will also come from the Sunyaev-Zeldovich (SZ) effect \citep{SunyaevZeldovich1970,SunyaevZeldovich1972,SunyaevZeldovich1980}, with future instruments like the Simons Observatory and CMB-S4 scheduled to come online over the next decade, as well as with the proposed high-resolution CMB-HD millimeter-wave survey. We note that even with current instruments like ALMA, a detection of the SZ effect may be possible \citep[see e.g.][]{Brownson2019}. Furthermore, the use of future radio observatories like the next-generation Very Large Array and Square Kilometre Array could potentially detect flux from forming quasars at even higher redshift \citep[e.g.][]{Latif2023}. Finally, the early growth of black holes themselves via mergers will be explored in the 2030s with the upcoming space-based gravitational wave observatory \textit{LISA} \citep{LISA}.

Given the wealth of observational resources incoming over the coming decades, a better theoretical understanding of how early, massive black holes affect galaxy evolution is therefore vital to make key predictions and help interpret the data from these instruments, and is what this paper aims to do.

This paper is organised as follows. In Section~\ref{Section:Methods} we introduce the numerical setup of our simulations, the models we use and the modifications we have made to encourage early, rapid black hole growth. In Section~\ref{Section:Results} we present our results, firstly of the black hole properties themselves and then of the galaxy and gaseous halo surrounding it. We discuss some of the caveats to our work in Section~\ref{Section:Discussion}, before summarising our conclusions in Section~\ref{Section:Conclusions}.

\section{Methods} \label{Section:Methods}
\subsection{Basic simulation properties}
The simulations in this paper use the cosmological, hydrodynamical code \textsc{arepo} \citep{Arepo}, that models gas on a moving mesh based on the Voronoi tessellation of a set of discrete mesh-generating points that can move with the flow. On-the-fly friends-of-friends \citep[FoF,][]{FOF} and \textsc{Subfind} halo finder \citep{Subfind1,Subfind2} algorithms are used to identify bound structures in the simulation output. 

This paper focuses on a zoom-in simulation of the largest halo in the Millennium box \citep{Millennium} at $z=6$, which has a virial mass $M_{200} = 6.9 \times 10^{12} \, \mathrm{M}_\odot$\footnote{$M_\mathrm{200}$ and the virial mass refer to the mass contained within $R_\mathrm{200}$, the radius within which the average density is $200$ times the critical density of the Universe at the redshift of interest.}. For further details of the setup of this zoom-in simulation see \citet{Sijacki2009}, though for reference the dark matter and gas mass resolutions of the simulation are $6.75\times10^6 \, h^{-1}$\,M$_\odot$ and $1.32\times10^6 \, h^{-1}$\,M$_\odot$, respectively, and the comoving softening lengths are 2.5\,kpc. This corresponds to mass-weighted mean cell radii of $\sim\!0.2\,$kpc at 2\,kpc from the halo centre at $z=6$, increasing to $\sim\!1\,$kpc at a distance of 50\,kpc. The protocluster goes on to form a galaxy cluster with halo mass $M_{200} = 4.6 \times 10^{15} \, \mathrm{M}_\odot$ at $z=0$. We have further simulated an additional five massive haloes at $z = 6$ from the Millennium simulation, with initial conditions taken from \citet{Costa2014}. The results of these simulations are shown in Appendix~\ref{AppA}. Like the Millennium simulation on which it is based, this work uses cosmological parameters $\Omega_\Lambda = 0.75$, $\Omega_\mathrm{m}=0.25$, $\Omega_\mathrm{b}=0.045$, $\sigma_8 = 0.9$ and $h = 0.73$, consistent with \textit{WMAP} cosmological constraints \citep{WMAP2007}.

\subsection{The \fable\ model} \label{Section:FABLE}

In this study we use variations atop the feedback model from the \fable\ suite of simulations \citep[described fully in][]{FABLE1}, itself a modified version of the feedback model from Illustris \citep{Illustris1,Illustris2,Illustris3,Illustris4}. The star formation and stellar feedback prescription for all runs remain unchanged from the base \fable\ model.

In the standard version of \fable, seed black holes of $10^5\, h^{-1} \, \mathrm{M}_\odot$ are placed at the gravitational potential minimum of haloes above a given halo mass threshold, $M_\mathrm{h, seed}$, which is an input parameter of the model (with a fiducial value of $5\times10^{10}\,h^{-1}$\,M$_\odot$). Black holes can grow through mergers with each other or through the accretion of gas, which is parameterised by a Bondi-Hoyle-Lyttleton-based model \citep{HoyleLyttleton1939,BondiHoyle1944,Bondi1952},
\begin{equation} \label{Eqn:MdotBondi}
    \dot{M}_\mathrm{Bondi} = \frac{4 \pi G^2 M^2_\mathrm{BH} \rho}{c_\mathrm{s}^3}\,,
\end{equation}  
where $G$ is the gravitational constant, $M_\mathrm{BH}$ is the black hole mass, and $\rho$ and $c_\mathrm{s}$ are the gas density and sound speed of the surrounding gas (the 32 nearest neighbour cells in the simulation). 

In reality, the medium a black hole sits in is likely to be denser than it is possible to probe with current cosmological simulations, both due to the limited range of spatial scales resolved and too-low interstellar medium densities modelled with sub-grid models for star formation. For this reason, the black hole accretion rate is often artificially boosted, and becomes $\dot{M}_\mathrm{BH} = \alpha \dot{M}_\mathrm{Bondi}$, where $\alpha = 100$ is the boost factor employed in \fable\ \citep[see e.g.][]{Springel2005}. 

The luminosity of an AGN (and therefore the energy potentially available for feedback) is related to the black hole accretion rate via $L = \epsilon_\mathrm{r} \dot{M} c^2$, where $\epsilon_\mathrm{r} = 0.1$ is the radiative efficiency of the black hole used in \fable. Due to the radiation output from the black hole, the accretion rate is assumed to be capped at the Eddington limit, which (assuming ionised hydrogen) is
\begin{equation} \label{Eqn:MdotEdd}
    \dot{M}_\mathrm{Edd} = \frac{4 \pi G M_\mathrm{BH} m_\mathrm{p}}{\epsilon_\mathrm{r} \sigma_\mathrm{T} c}\,,
\end{equation}
where $m_\mathrm{p}$ is the proton mass, $c$ is the speed of light and $\sigma_\mathrm{T}$ is the Thomson cross-section. The accretion rate is further modulated by a `pressure criterion', which reduces the accretion rate during epochs where a large black hole is embedded in low-density background gas, as this was found to sometimes lead to the unphysical inflation of hot, diffuse bubbles due to feedback \citep[see section 2.6.2 of][]{Illustris1}.

\fable, like Illustris, uses a dual-mode feedback model for AGN, depending on the black hole accretion rate. For the quasar mode, at high accretion rates, 10 per cent of the feedback energy is isotropically injected as thermal energy to neighbouring cells of the black hole. In the original Illustris, this is done continuously, which can make the feedback less effective as energy is immediately radiated away. In \fable\, a duty cycle is imposed - specifically, feedback energy is stored for 25\,Myr before being injected at once \citep[motivated by the AGN model introduced by][]{BoothSchaye2009}. This can make quasar-mode feedback more effective. At low accretion rates a radio mode is invoked, which injects hot bubbles into the gas around the black hole \citep[following][]{Sijacki2007} when the black hole mass changes by 1 per cent. 

Dynamical friction cannot be well resolved in these simulations, so black holes are migrated to (and kept in) the centre of their host halo by repositioning them to be at the minimum gravitational potential of particles within their smoothing length at each active timestep \citep[as per][]{Springel2005}. This process occurs much faster than it would under dynamical friction. In addition, mergers happen immediately when two black hole particles are within each other's smoothing lengths. The combination of these effects means that mergers are too efficient in simulations with such models, like Illustris \citep{Illustris1}, IllustrisTNG \citep{Weinberger2017a} and \fable\ \citep{FABLE1}. 

\subsection{Modifications} \label{Section:Modifications}

\begin{table}
    \centering
    \caption[Summary of the parameters used in each simulation in this paper.]{A summary of the three simulations we have performed and the model parameters we have used for each run. The columns list the run name, the halo mass threshold in which seed black holes are placed, the maximum allowed Eddington fraction, and the coupling strength of quasar mode feedback (see Section~\ref{Section:FABLE}).}    
    \begin{tabular}{c|c|c|c}
        
        \hline
    
        Run name & $M_\mathrm{h, seed}$ [$\mathrm{M}_\odot$] & $f_\mathrm{Edd, max}$ & Feedback coupling strength \\
          
        \hline

        NoAGN & - & - & - \\
        FABLE & $5\times10^{10}$ & $1$ & $0.1$\\
        Reference & $1\times10^{9}$ & $2$ & $0.05$\\

        \hline
    \end{tabular}

    \label{tab:Runs}
\end{table}

We present results from three zoom-in simulations in this paper, with the relevant parameters of each run described in Table~\ref{tab:Runs}. The base \fable\ run (stylised as ``FABLE'' in the figures) uses the standard \fable\ model without modification, as a benchmark simulation to compare to. There is also an equivalent ``NoAGN'' run, which uses the standard \fable\ setup except with all black hole growth and feedback switched off.

For the modified run (henceforth termed ``Reference'' runs) we make a number of changes to the base \fable\ model to promote black hole growth. We switch off the duty cycle for quasar feedback implemented in \fable, returning to a continuous injection of feedback energy that is less efficient at driving outflows. The low accretion rate ``radio'' mode of feedback is also switched off for the same reason, though we note this mode may not be dominant at high redshifts (see Fig.~\ref{fig:BHEddFracfull}). Additionally, we switch off the pressure criterion present in the original Illustris simulations, described above, though we note this has little effect.

We seed black holes earlier (and also generate more of them), by lowering the halo mass in which black holes are seeded to $M_\mathrm{h, seed} = 10^9\,h^{-1} \,\mathrm{M}_\odot$. This increases the redshift at which the first black hole is seeded from $z\sim13$ to $z\sim18$, gaining an extra $\sim\!100$\,Myr for the black hole to grow. Reducing the halo mass threshold for black hole seeding has the additional effect of significantly increasing the number of black hole particles in the simulation, as also noted by \citet{Huang2020} and \citet{Bhowmick2022}. The vast majority of the final black hole mass comes from gas accretion onto the main black hole and its progenitors, though we note at early times black hole seed mergers contribute a higher fraction of the mass, with mergers driving the majority of the main progenitor growth above $z\sim10$ in the Reference model \citep[see also][]{Sijacki2009,Bhowmick2022}. It remains an open question if this growth channel can be so efficient, however to investigate this further would require more comprehensive black hole dynamical friction and merger modelling and is beyond the scope of this work. In our simulations this effectively `boosts' the black hole mass at very high redshifts, which then facilitates faster growth from gas accretion due to the Bondi accretion model's quadratic dependence on black hole mass (Equation~\ref{Eqn:MdotBondi}). As also discussed in \citet{Sijacki2009}, this can affect the growth history of the black hole, but does not significantly affect the final black hole mass.

We also switch off the basic, proximity-based AGN radiation field present in \fable, Illustris, and IllustrisTNG, which exists as a superimposed radiation field on top of the uniform UV background \citep[and does not self-consistently follow AGN radiation with radiative transfer, see][for details]{Illustris1}. We note that the radiation fields of the increased number of growing black holes mentioned above does have an impact on the thermal state of the IGM outside of the halo accretion shock, which could impact accretion through the CGM. However to investigate this properly would require on-the-fly radiative transfer simulations that are beyond the scope of this work. We find that turning this radiation field off makes little difference to the properties of the black hole or central galaxy, so to be consistent between the runs with and without black holes we leave this off in this work. 

Importantly, we also reduce the efficiency of feedback energy being injected by the quasar-mode model, from 10 per cent (as in \fable\ and TNG) to 5 per cent (as in the original Illustris). We note that this does \textit{not} mean that the total injected feedback energy is reduced, as this also crucially depends on the black hole mass, which as we will show is higher. The main effect of reduced feedback coupling efficiency is to promote earlier growth of black holes, with feedback-regulated growth kicking in at a somewhat higher black hole mass. We also allow the black hole to accrete gas at mildly super-Eddington rates, with a maximum-allowed Eddington fraction $f_\mathrm{Edd, max} = 2$. These changes have the dual effect of reducing the effectiveness of AGN feedback earlier in the evolution of the halo, allowing more gas to remain present in the central regions of the forming galaxy, as well as allowing more of that gas to be consumed by the black hole. 

We note that the purpose of this paper is not to find the `correct' way of growing a `gargantuan' black hole, nor do we claim the changes we make are the only way to do so. Our aim is to show a \textit{plausible} pathway to grow a black hole with a mass in agreement with the brightest high-redshift quasars, to explore the impact that feedback from such an object could have.

\subsection{Diagnostics}

In addition to the properties of the black holes themselves, we also investigate the impact of the quasar on the galaxy and its surroundings via a number of observables accessible to current or future instruments. 

\subsubsection{Obscuration and observability} \label{Section:Obscuration}

We track the bolometric luminosity of accreting black holes on-the-fly in the simulations, which we can compare to observed samples of high-redshift quasars. However, such a comparison is not trivial. Most high-redshift quasars discovered to date were found in the rest frame UV by wide area surveys like SDSS. Significant amounts of gas and dust in the quasar host galaxy along the line of sight to Earth can drastically reduce the apparent brightness of quasars in the UV. To perfectly model this obscuration would involve radiative transfer simulations and a model for dust, which is beyond the scope of this paper, however we can make relatively simple approximations for both that can allow for better comparison to observations. 

To start with, we calculate the bolometric luminosity of the quasar from the simulation output,
\begin{equation} \label{Eqn:Lbol}
    L_\mathrm{bol} = \epsilon_\mathrm{r} \dot{M}_\mathrm{BH} c^2,
\end{equation}
where $c$ is the speed of light, $\epsilon_\mathrm{r} = 0.1$ is the radiative efficiency of the black hole in \textsc{arepo} and $\dot{M}_\mathrm{BH}$ is calculated from the boosted Bondi-Hoyle-Lyttleton rate as described in Section~\ref{Section:FABLE}. To then estimate the UV luminosity we perform a reverse bolometric correction using the fit of \citet{Venemans2016},
\begin{equation}
    \log \left( \frac{L_\mathrm{bol}}{10^{46}\, \mathrm{erg}\, \mathrm{s}^{-1}} \right) = 0.459 + 0.911 \times \log \left(\frac{\lambda L_\mathrm{\lambda}(1450\textrm{\AA})}{10^{46}\, \mathrm{erg}\, \mathrm{s}^{-1}} \right).
\end{equation}
We then calculate the intrinsic (unattenuated) quasar UV flux at the cosmological distance to the redshift of the snapshot. This flux would then be attenuated by dust in the quasar host galaxy,
\begin{equation}
    F_\mathrm{obs} = F_\mathrm{int} e^{-\tau_\lambda},
\end{equation}
where $\tau_\lambda$ is the UV optical depth, defined as
\begin{equation}
    \tau_\lambda = \kappa_\lambda \frac{f_\mathrm{DG}}{f_\mathrm{DG, MW}} \Sigma_\mathrm{gas} = \kappa_\lambda \frac{1}{f_\mathrm{DG, MW}} \Sigma_\mathrm{dust}. 
\end{equation}
Here, $\kappa_\lambda$ is the UV extinction cross-section, which we take in this work to be 1000\,cm$^2$\,g$^{-1}$ \citep[following][]{Thompson2015}, valid for a Milky Way dust-to-gas ratio $f_\mathrm{DG, MW}$. $\Sigma_\mathrm{gas}$ is the total gas column density, and $\Sigma_\mathrm{dust} = f_\mathrm{DG} \Sigma_\mathrm{gas}$ is the dust column density. We note that we calculate the dust column density from the cold metal mass, as described below, rather than from a dust-to-gas ratio applied to all gas.

We do not track dust on-the-fly in the simulation, so we make the simple approximation that dust makes up 15 per cent of the mass stored in metals \citep[as in][]{DiMascia2021,Vito2022}. We therefore calculate the dust mass of a quasar host by summing up 15 per cent of the cool metal mass (with $T<5\times10^4\,\mathrm{K}$, to approximate the effect of thermal sputtering) within 1.5\,kpc of the black hole at each timestep, with the radius chosen to match the largest observed size of the dust continuum observed by ALMA in the sample of \citet{Gilli2022}. We note that choosing a larger radius does not change our results, as the dominant contribution to obscuration comes from gas close to the centre of the galaxy. We find our results on obscuration do not strongly depend on the 15 per cent dust to metal ratio parameter, and that our dust masses are in reasonable agreement with observational samples (see Section~\ref{Section:ObscurationResults}). 

To calculate the column density, we generate $\sim\!10,000$ radial rays, of length 5\,kpc, from the position of the black hole at each snapshot, equally spaced on a sphere. A ray length of 5\,kpc was chosen as it always exceeds twice the stellar half-mass radius of the central galaxy and so gives us an indication of the host absorption. We also repeated the calculations with rays of up to 50\,kpc and found our results did not change significantly, with only a slight decrease in the number of rays with the lowest column densities when more dense gas in the CGM was included, similar to what was found by \citet{Ni2020}. Each ray contains 1024 logarithmically spaced radial points, which we associate with the properties (density and metallicity) of their nearest cell. By numerically integrating along each radial ray (assuming values are piecewise-constant along the ray), we therefore calculate a distribution of the column densities of hydrogen ($N_\mathrm{H}$) and metals at every snapshot of the simulation. 

After adding these steps together we reach an observed flux, which we then convert into an apparent magnitude to be able to compare to future observational expectations. 

\subsubsection{H\textsc{i}} \label{Section:HIMethods}

The dense gas in and around the quasar host could often be self-shielded from background radiation fields, however a full treatment of this would require radiative transfer. For the purposes of this paper, we therefore make approximations to estimate the H\textsc{i} mass of each cell. Specifically, we follow the prescription of \citet{Bird2013}, where gas is assumed to transition from self-shielded at high densities to being in equilibrium with the UV background at low densities \citep[see equation 1 in][]{Bird2013}. 

\subsubsection{Ly$\alpha$} \label{Section:LyaMethods}

An important observational probe of the cool gas content and its dynamical state is that of the bright, resonant Ly$\alpha$ line, which can be traced using MUSE of \textit{JWST} observations. A full treatment of Ly$\alpha$ luminosities would require detailed photoionisation and radiative transfer calculations, taking into account both stellar and AGN radiation, which we leave to future work. However, we can again make approximations to estimate the luminosity and morphology from our existing simulations. We consider the Ly$\alpha$ luminosity that comes from collisionally ionised hydrogen, using the temperature dependent rate coefficients from appendix A of \citet{SmithA2022}. Ly$\alpha$ photons are emitted from collisionally ionised hydrogen within a fairly narrow temperature range ($3\times10^4\, \mathrm{K} \lesssim T \lesssim 5\times10^4$\,K), and we note we exclude all star forming gas from these calculations as its temperature is not self-consistently tracked.  

We note that our simplistic choice of Ly$\alpha$ production means we are excluding the production of Ly$\alpha$ from radiative recombination, and the absorption of Ly$\alpha$ photons by dust. We also neglect the contribution of Ly$\alpha$ photons from the broad line region of the AGN itself, which can be an important source of photons to power a Ly$\alpha$ nebula \citep[see][]{Costa2022}. Due to these reasons, the total Ly$\alpha$ brightness of our haloes should be treated with caution. However, as a proof of concept, we can still investigate how the brightness and morphology \textit{changes} between different feedback models. 

\subsubsection{Hot gas definition} 
For some of the properties of the gaseous halo, such as X-ray emission and the thermal SZ effect, only cells containing hot gas should contribute. In this work, this hot gas is determined by removing cold clumps from the simulation output in the same way as \citet{Bennett2020,Bennett2022}, using a rescaled version of the method described in \citet{Rasia2012}. We separate the `hot' gas phase from the `cool' using the condition 
\begin{equation} \label{Eqn:Rasia12Cut}
    T > N \times \rho^{0.25}\,,
\end{equation}
where $T$ and $\rho$ are the temperature and density of the gas and $N$ is a normalisation factor. We assume the same fixed normalisation factor $N = 3\times10^6$~keV cm$^{3/4}$~g$^{-1/4}$ as \citet{Rasia2012}, where temperature is in keV and density is in g~cm$^{-3}$. We then rescale this relation by the temperature, $T_{500}$, of the simulated halo, relative to the mean $T_{\mathrm{500}}$ of the haloes in \citet{Rasia2012}. While this cut was developed using simulations of low-redshift galaxy clusters, we have verified it still separates the hot phase of the CGM of our haloes at high redshift.

\subsubsection{X-ray} \label{Section:XrayMethods}

We produce mock X-ray images of the halo in a number of ways. Firstly, we use the simple Bremsstrahlung approximation \citep[as in e.g.][]{Bennett2022}
\begin{equation}
    L_\mathrm{X, Br} = \frac{1.2\times10^{-24}}{\mu^2 m_\mathrm{p}^2} \sum_i m_i \rho_i \mu_i^{-2} T_i^{1/2}\,,
\end{equation}
where $m$, $\rho$, $\mu$ and $T$ are the mass, density, mean molecular weight and temperature, respectively, of individual hot gas cells. 

We then produce more realistic X-ray mocks to make predictions for several planned or proposed X-ray telescopes. Firstly, using the \texttt{pyXSIM} package \citep{ZuHone2016}, itself based on the \textsc{phox} \citep{Biffi2012,Biffi2013} and \texttt{yt} \citep{yt} packages, we generate and project X-ray photons from the simulation output. Specifically, we take a spherical region of radius 500\,kpc around the centre of the largest halo at $z=6$ and then generate thermally emitted photons from the hot gas component assuming collisional ionisation equilibrium. We note that photoionisation and resonant scattering could also play a significant role in the gaseous halo of these quasars, however the modelling of this in \texttt{pyXSIM} is not currently designed for such high redshift and so we leave this explicit modelling to future work. Importantly, this means that the X-ray brightness of the halo presented in this work is a somewhat conservative lower limit. We further note that to minimize the impact of a recent burst of quasar feedback at $z=6$ in the very central regions, when making our mock X-ray images we do not include gas in the central 1\,kpc, allowing us to focus on the integrated quasar feedback effect on the hot halo. 

After the photon list is generated, we use the \texttt{SOXS} \citep{ZuHone2023} package to produce mock 1\,Ms observations using the proposed \textit{ATHENA} \citep{ATHENA}, \textit{AXIS} \citep{AXIS}, and \textit{Lynx} \citep{Lynx} X-ray telescopes, including astrophysical foregrounds and instrumental backgrounds. We produce images that include the whole (planned) bandpass of each instrument, which varies slightly between them, though we note using the same energy limits for the three mock observations does not change our results. We note that while other proposed X-ray telescopes such as \textit{XRISM} \citep{XRISM} or the \textit{Line Emission Mapper} \citep{LEM} would not be able to spatially resolve the gaseous halo of the quasar, their high-resolution spectra could allow interesting down-the-barrel studies of the halo using the light of the quasar itself.

\subsubsection{SZ effect}

We also produce mock maps of the SZ effect along a line-of-sight (chosen to be along the $+z$ axis of the simulation box in this work). For the thermal SZ (tSZ) effect, the projected Compton-$y$ value along a line-of-sight is proportional to the integrated electron pressure,
\begin{equation}
    y = \frac{\sigma_\mathrm{T}}{m_\mathrm{e} c^2} \int P_\mathrm{e} \, \mathrm{d}z\,, 
\end{equation}
where $\sigma_\mathrm{T}$ is the Thomson scattering cross-section, $m_\mathrm{e}$ is the electron mass, $c$ is the speed of light and $P_\mathrm{e} = n_\mathrm{e} k_\mathrm{B} T$ is the electron pressure. We also make the assumption that the electron temperature $T_\mathrm{e}$ is the same as the gas temperature, $T$. The amplitude of the kinetic SZ effect (kSZ), $w$, is proportional to the line-of-sight gas velocity, $v_\mathrm{z}$,
\begin{equation}
    w = -\frac{\sigma_\mathrm{T}}{c} \int n_\mathrm{e} v_\mathrm{z} \, \mathrm{d}z\,,
\end{equation}
and can provide additional constraints on the dynamics of gas in haloes. The kSZ effect can come from any ionised gas moving with sufficient bulk velocity, even at lower temperatures than the hot halo, unlike X-ray emission and the tSZ effect. We therefore create these maps using all ionised gas that is not star-forming and that has a temperature $T > 3\times10^4$\,K.

\section{Results} \label{Section:Results}

\begin{table*}
    \centering
    \caption[Summary of the final black hole and stellar masses in the simulations of this paper.]{A summary of the $z=6$ black hole mass in each of the zoom-in simulations in this work, as well as the stellar mass and instantaneous star formation rate of its host galaxy, evaluated within twice the stellar half-mass radius. We also report the value of the stellar half-mass radius, $R_*$, and the one-dimensional stellar velocity dispersion of stars within the stellar half-mass radius, $\sigma_\mathrm{*,1D} = \sigma_\mathrm{*,3D} / \sqrt{3}$.}    
    \begin{tabular}{c|c|c|c|c|c}
        
        \hline
    
        Run name &  $M_\mathrm{BH}$ [$\mathrm{M}_\odot$] & $M_\mathrm{*}$ [$\mathrm{M}_\odot$] & SFR [$\mathrm{M}_\odot \, \mathrm{yr}^{-1}$] & $R_*$ [kpc] & $\sigma_\mathrm{*,1D}$ [km\,s$^{-1}$]\\
          
        \hline
        NoAGN & - & $3.5\times10^{11}$ & 1255 & 0.53 & 426\\
        FABLE & $1.4\times10^{9}$ & $2.8\times10^{11}$ & 306 & 0.68 & 406\\
        Reference & $1.3\times10^{10}$ & $1.5\times10^{11}$ & 210 & 1.78 & 353\\
        
        \hline
    \end{tabular}

    \label{tab:FinalValues}
\end{table*}

In Table~\ref{tab:FinalValues} we show the $z=6$ black hole mass, as well as the stellar mass, star formation rate (SFR), stellar half-mass radius, and stellar velocity dispersion of our the quasar host galaxy for all of the simulations we have performed. With the changes we made to \fable\ to generate our Reference run, we have a final black hole mass in excess of $10^{10}\,\mathrm{M}_\odot$ by $z=6$, an order of magnitude larger than in \fable. Despite this, the drop in its host galaxy's stellar mass and SFR are less dramatic, with a decrease of $\sim\!\!50$ and $\sim\!\!33$ per cent, respectively. Interestingly, the stellar half-mass radius more than doubles in the Reference run, indicating that the strong AGN feedback not only reduces the stellar mass but also makes the quasar host galaxy significantly less compact. We investigate this further in Section~\ref{Section:StellarStructure}. Unsurprisingly, the less compact and less massive quasar host galaxies in the Reference run results in a lower stellar velocity dispersion, and furthermore galaxy morphology becomes somewhat more `disky'. 

\subsection{Black hole mass evolution} \label{Section:BHMass}

\begin{figure}
    \centering
    \includegraphics[width=\linewidth]{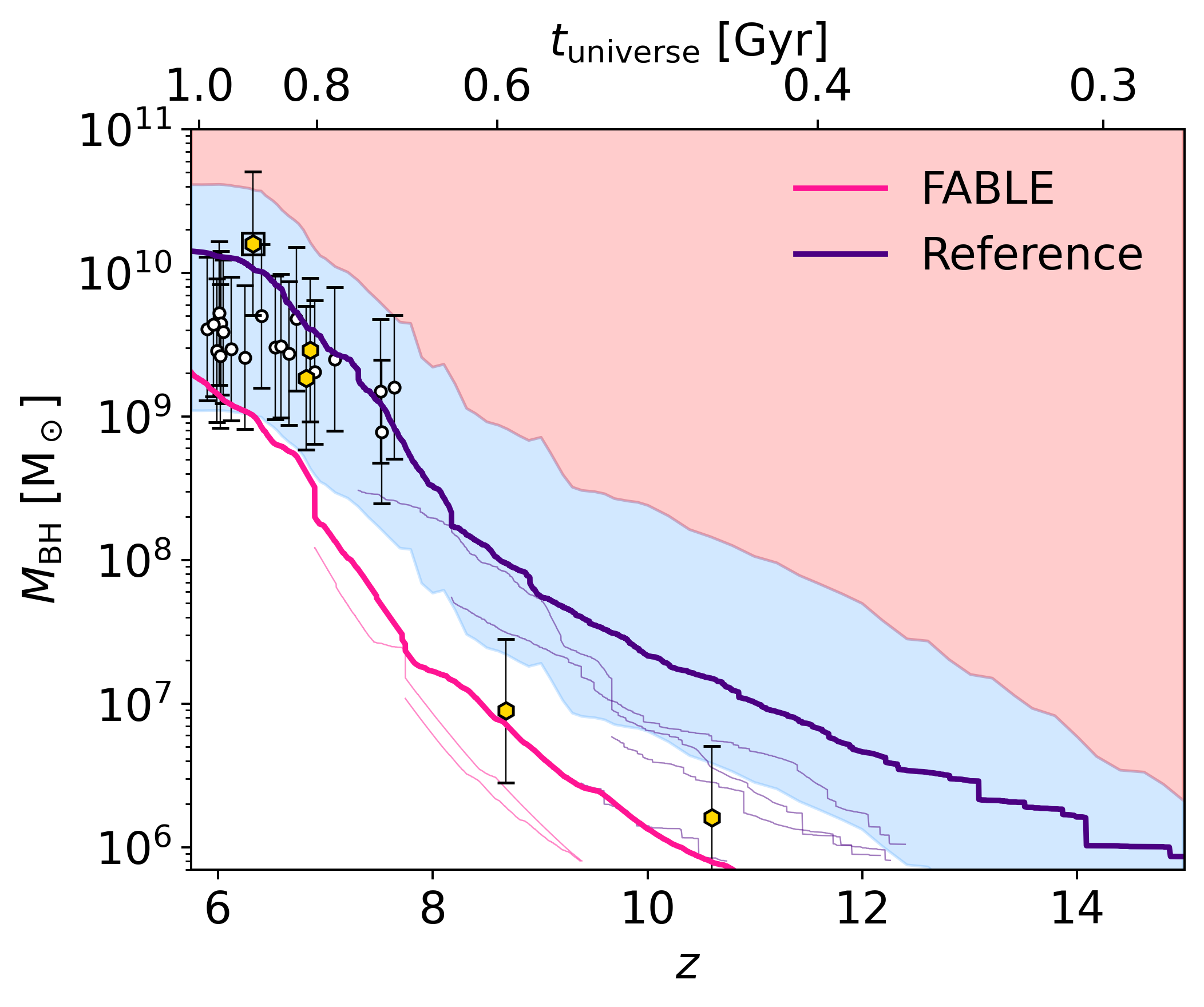}
    \caption[Black hole mass growth curve for the different simulations.]{Growth of black hole mass as a function of redshift for the largest black hole in each run considered in this paper. Thinner lines show the mass growth of black holes that merge into the primary progenitor in each simulation. The blue shaded area shows the region at which a black hole would exceed the median $z=0$ $M_\mathrm{BH} / M_*$ relation, and the red area shows the equivalent for a ratio of $M_\mathrm{BH} / M_* = 0.15$ (the current record $z=0$ ratio). The calculation of these regions use the stellar mass from \fable\ only for clarity. Individual points show where observations of black holes lie: open black circles show masses $>2\times10^{9}\,$M$_\odot$ estimated from the Mg\textsc{ii} line widths \citep{KimIm2019,Yang2020,Wang2021}, gold hexagons show data from \textit{JWST}, all using H$\beta$ line widths \citep{Eilers2023,Marshall2023,Larson2023} apart from the Mg\textsc{ii} measurement of a black hole in GN-z11 at $z=10.6$ \citep{Maiolino2023a}. Error bars show a representative $0.5$\,dex error, from the systematic uncertainty in estimating virial black hole masses from observed line widths \citep[e.g.][]{McLureDunlop2004,VestergaardOsmer2009}. The black square highlights the data from J0100+2802, the brightest observed high-redshift quasar. The changes we made in our Reference simulation allow us to grow a black hole that matches both the highest redshift observed quasars \textit{and} the brightest known high-redshift quasar.}
    \label{fig:BHgrowth}
\end{figure}

In Fig.~\ref{fig:BHgrowth} we show the mass growth curves of the most massive black hole in our simulations, compared to the largest known high-redshift black holes from observations. Observed black hole masses \citep[estimated from the virial theorem applied to the broad-line region of an AGN, e.g.][]{McLureJarvis2002,VestergaardOsmer2009} are shown from a range of recent works as individual points. These data are primarily taken from Table 3 of \citet{KimIm2019}, who estimate black hole masses, luminosities, and star formation rates of quasars and their hosts from a selection of literature data  \citep{Wang2007,Wang2008,Wang2011,Wang2013,Wang2016,Willott2010,DeRosa2011,Venemans2012,Venemans2017,Omont2013,Banados2015,Banados2018,Wu2015,Mazzucchelli2017,Decarli2018,Eilers2018,Shen2019}. In this and subsequent figure, we only include data from \citet{KimIm2019} for black holes with an inferred mass $>2\times10^9\,$M$_\odot$, both for visibility on Fig.~\ref{fig:BHgrowth} and because the focus of this paper is on the largest black holes at high redshift. 

We also include data from two very high-redshift quasars discovered since \citet{KimIm2019} was published \citep{Yang2020,Wang2021} in Fig.~\ref{fig:BHgrowth}. All of these use the Mg\textsc{ii} line width to estimate black hole masses. In addition, we show four measurements of accreting black hole masses inferred from H$\beta$ line widths, made using \textit{JWST} \citep{Eilers2023,Marshall2023,Larson2023}. Finally, we include the recent measurement of a black hole in GN-z11, inferred using Mg\textsc{ii} line widths observed with \textit{JWST} \citep{Maiolino2023a}. In this and subsequent plots, we use a black square to highlight the available data for J0100+2802, the largest known high-redshift quasar, whose `gargantuan' inferred black hole mass of $\sim\!10^{10}\,\mathrm{M}_\odot$ strongly motivated this project.

The blue shaded region in Fig.~\ref{fig:BHgrowth} shows where a black hole mass would exceed the $z=0$ median black hole to stellar mass ratio \citep[$\sim\!0.4$ per cent, e.g.][]{KormendyHo2013} for the stellar mass in the \fable\ run. The red region shows the equivalent for a black hole to stellar mass ratio equal to the current claimed maximum $M_\mathrm{BH} / M_*$ ratio \citep[$\sim\!15$ per cent,][]{vandenBosch2012,Seth2014}.

The pink line in Fig.~\ref{fig:BHgrowth} shows the largest black hole in the \fable\ run, which undergoes approximate exponential growth after $z=10$ and reaches a mass at $z=6$ of $1.4\times10^9\,\mathrm{M}_\odot$. This is consistent with the observed masses of some quasars at this epoch, suggesting the fiducial \fable\ model can potentially produce of the less-massive observed objects at $z=6$. We note that the two earliest discovered candidate accreting black holes to date, with inferred masses of $\sim\!10^7$\,M$_\odot$ at $z=8.7$ \citep{Larson2023} and $\sim\!2\times10^6$\,M$_\odot$ at $z=10.6$ \citep{Maiolino2023a}, both lie fairly close to the \fable\ growth path. This implies that even in an Eddington-limited scenario with active feedback, such a black hole could easily be a progenitor of $\sim\!10^9$\,M$_\odot$ black holes powering quasars by $z\sim6$. However, our results imply that these two very high-redshift black holes are less likely to be progenitors of the brightest discovered high-redshift quasars. The further implication is that either there could be a population of even more massive black holes at $z>8$ that have yet to be discovered, or that significant super-Eddington accretion is widespread and, crucially, \textit{sustained}, for such early black holes to grow into `gargantuan' ones a short time later.

Considering both the largest known high-redshift black hole ($\sim\!10^{10}\,\mathrm{M}_\odot$ at $z=6.3$) and those discovered at even higher redshift ($z\gtrsim7.5$), the \fable\ model only produces black holes an order of magnitude below the observed mass estimates of these brightest quasars. This implies that either the fiducial \fable\ model simply cannot produce such massive black holes - perhaps unsurprising given \fable\ has been in no way tuned to reproduce the properties of high-redshift quasars - or an even rarer host halo might be required. 

Our halo is already the most massive in the entire Millennium simulation box (with $500 \, h^{-1}$ comoving Mpc on a side), suggesting significantly bigger boxes would be needed to find a more massive halo. However, it is worth remembering that such haloes ($\gtrsim10^{13}\,$M$_\odot$) are likely to be very rare at high redshifts in a $\Lambda$CDM Universe. In fact, in the MillenniumXXL simulation \citep{Angulo2012}, which is $4.1 \, h^{-1}$ comoving Gpc on a side, the most massive FoF halo\footnote{Note that $M_{\rm FoF}$ and $M_{\rm 200}$ are distinct measures, with the former often being larger.} has $\sim 4\times10^{13} \,$M$_\odot$. Rescaling MillenniumXXL to \textit{Planck}-like cosmology would lead to the most massive FoF halo mass $M_\mathrm{FoF} \sim 2.7\times10^{13} \,$M$_\odot$ \citep[see figure 2 of][]{Angulo2012}, implying that our relatively high value of $\sigma_{\rm 8}$ somewhat helps us simulate a rarer proto-cluster overdensity, which would be more readily found in a simulated volume larger than that of Millennium.

When we move to our Reference run (the dark purple line in Fig.~\ref{fig:BHgrowth}) we see how the changes we have made to the \fable\ model have a significant effect. Firstly, the black hole grows much earlier due to the higher seeding redshift, reaching the same mass as the $z=6$ \fable\ run at $z\sim7.5$. The final black hole mass of the Reference run at $z=6$ is consistent with the largest observed black holes at this epoch, an order of magnitude above the fiducial \fable\ run. The Reference run is also consistent with the brightest quasars at $z>7$, suggesting these even higher redshift black holes may be on a track to becoming $10^{10}$\,M$_\odot$ black holes by $z\sim6$.

One may envisage that the changes we made to the \fable\ model could overproduce the number density of bright quasars as we are extracting our halo from the Millennium simulation, whose volume is about hundred times smaller than that of surveys probing high redshift quasars (but see discussion above about the effects of cosmology). It is important to stress here that the host halo mass and its local overdensity are not the sole requirements for successful growth of very massive black holes, as the early assembly history of these rare massive haloes plays a significant role \citep[for further discussion see][]{Costa2014}. Nevertheless, to better explore the imprint of our changes with respect to the \fable\ model, we have simulated additional five most massive haloes from the Millennium simulation, taking the initial conditions from \citet{Costa2014}. The growth curves for the most massive black hole in each of these haloes both for the \fable\ and the Reference setup are shown in Appendix~\ref{AppA}. Reassuringly, we see that only in the most massive Millennium halo black hole mass of $\sim 10^{10}$\,M$_\odot$ is reached by $z \sim 6$ for the Reference setup. The black hole masses in the other five halos are consistent with observationally-estimated quasar masses, while the \fable\ setup produces black holes which are typically one order of magnitude less massive. Excitingly, upcoming \textit{JWST} observations together with theoretical models exploring even rarer objects, may soon help us distinguish between these different growth models \cite[see e.g.][]{Jeon2023}.  

Focusing now on the co-evolution between black holes and their host galaxies, we note that the most massive black hole in the \fable\ run lies below the median $z=0$ black hole to stellar mass ratio (the bottom of the blue shaded region) for most of its evolution but falls onto the relation by $z=6$.
In the Reference run, the black hole lies significantly above the median $z=0$ relation throughout, but never exceeds the maximum known ratio at low redshift (15 per cent). We note that a black hole following the median $z=0$ mass ratio to high redshift is not required, or even necessarily expected, and the evolution of this ratio is both observationally and theoretically poorly constrained \citep[see e.g. results and discussion in][]{Habouzit2022b}. 

At earlier times in both the \fable\ and Reference runs, several major mergers cause jumps in the black hole mass, with the merging black holes (that have a mass above $10^{7}\,\mathrm{M}_\odot$) shown as thinner lines in Fig.~\ref{fig:BHgrowth}. In contrast to the work of \citet{Zhu2022}, we find that major black hole-black hole mergers can play a significant role in their mass growth at certain points in the halo's evolution, and may help black holes grow to very large masses in the early Universe \citep[see also][]{Sijacki2009}. 

There is also expected to be a range of black hole seed masses in the early Universe. While our fiducial and Reference models both use a seed of $10^5$\,M$_\odot$, we note that an even larger seed with a rapid early growth could allow such an extreme object at an even higher redshift. Recent work has also suggested that mergers between massive haloes can lead directly to even larger seed mass black holes \citep{Mayer2023}. Better-constrained quasar luminosity functions pushing to higher redshift will be needed to know how realistic such paths could be. In addition to this, super-Eddington accretion beyond twice the Eddington limit could also play a role. However, this can also increase the amount of inputted feedback energy, and so it is unclear what the effective change in total black hole mass would be \citep[see e.g.][]{Zhu2022}.

\subsection{Accretion, obscuration and observation} \label{Section:ObscurationResults}

\begin{figure*}
    \centering
    \includegraphics[width=0.49\linewidth]{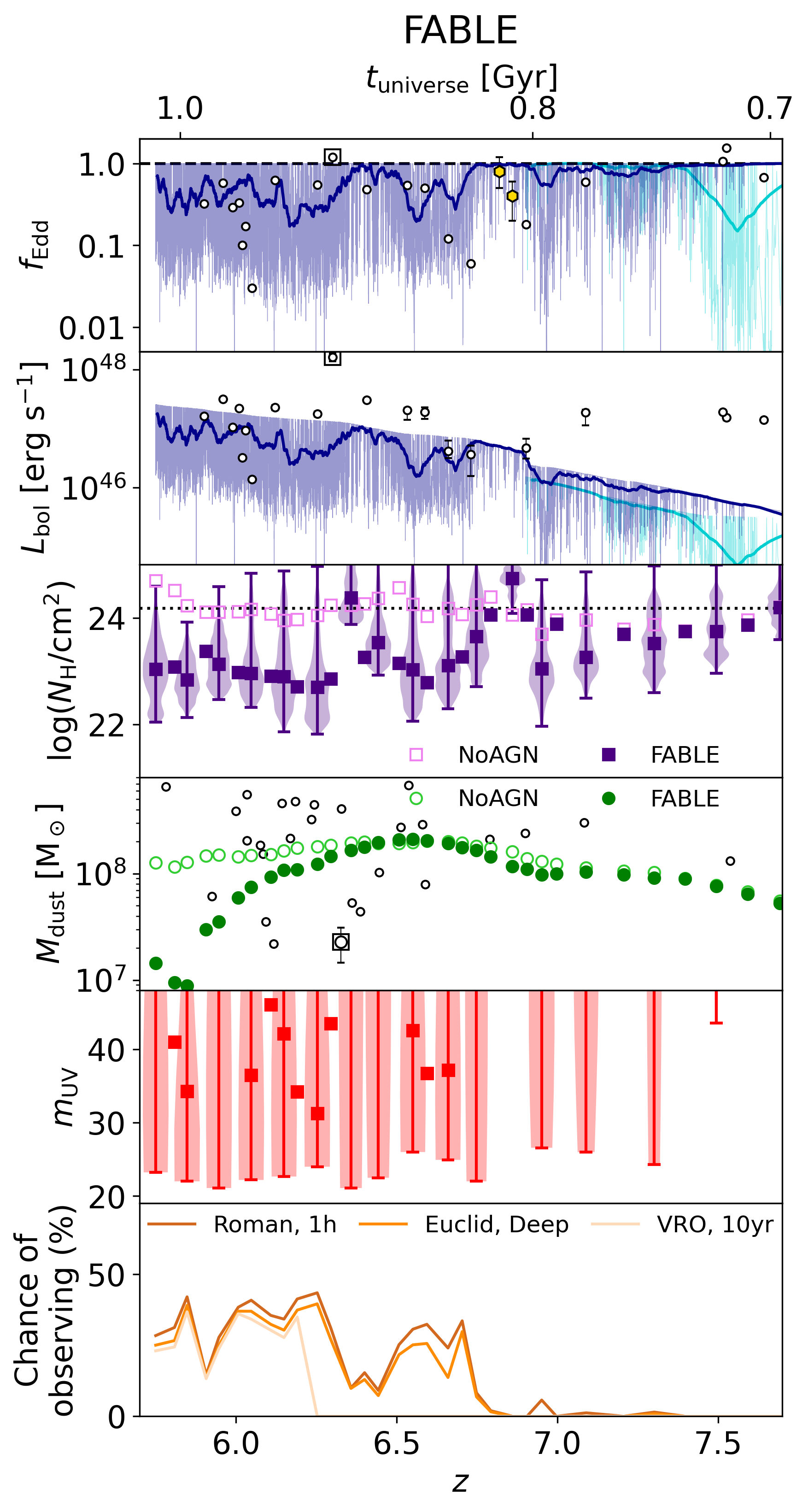}
    \includegraphics[width=0.49\linewidth]{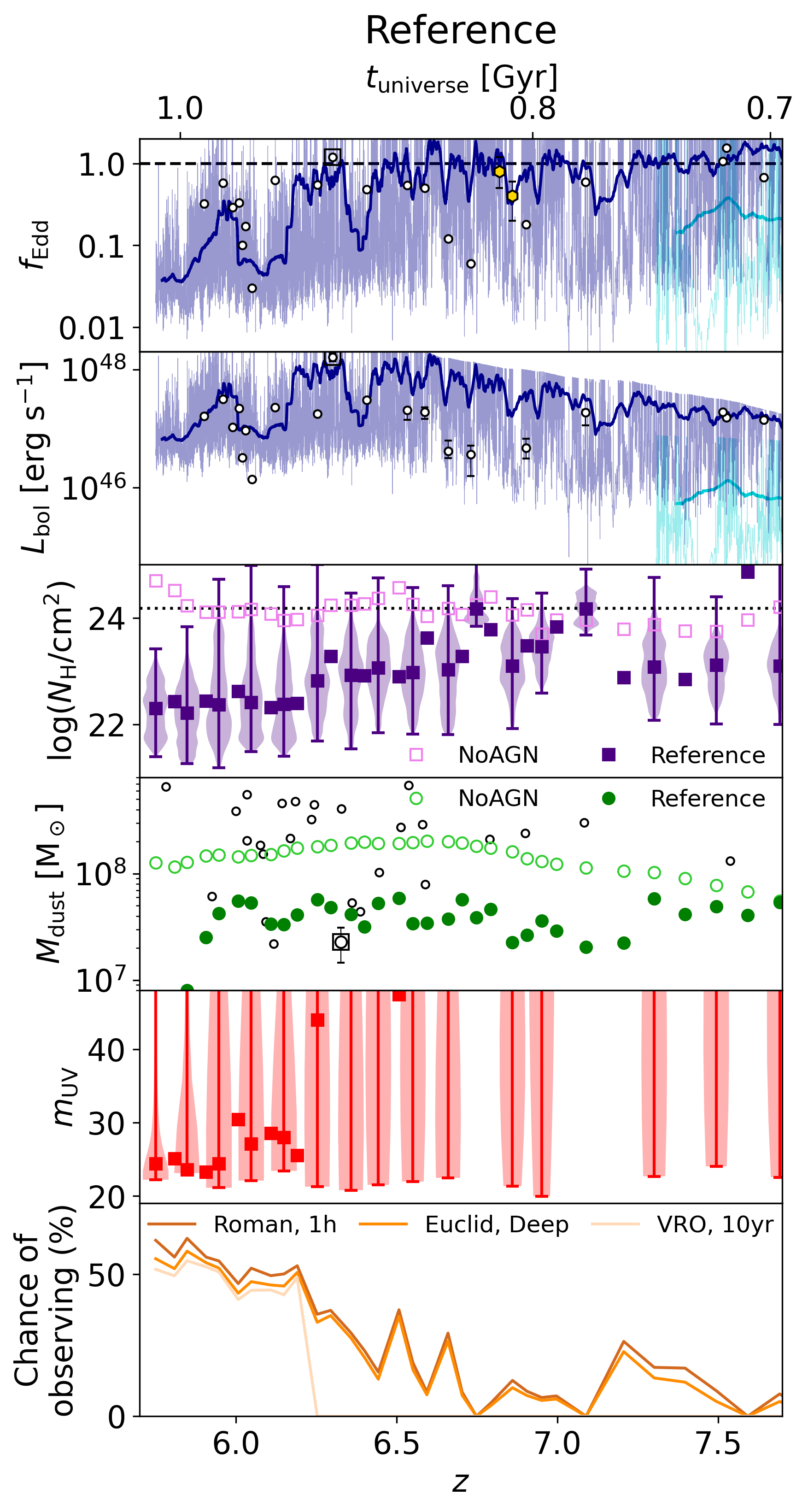}    
    \caption[Evolution of black hole observability properties in the \fable\ and Reference simulations.]{Evolution of black hole accretion and observability properties for the \fable\ (left) and Reference (right) simulations. 
    Eddington fraction (\textit{top row}) and bolometric luminosity (\textit{second row}) are shown for the largest black hole and its largest progenitor. Faded lines show the value at every timestep the black hole particle is active, and the darker lines show a rolling mean with a width of 100 timesteps. Observational data points are shown as either open circles or gold hexagons (for data from \textit{JWST)}, with errors where available, from the sources cited in Section~\ref{Section:BHMass}. The dashed horizontal line shows the Eddington limit, $f_\mathrm{Edd} = 1$. The black squares highlight the data from J0100+2802, the largest observed high-redshift quasar. The \fable\ cannot grow black holes big enough to explain the luminosity of very bright, high-redshift quasars, but our Reference run can.
    \textit{Third row}: distribution of hydrogen column densities along $\sim\!10,000$ rays of length 5\,kpc from the black hole. Median values are shown as solid squares for each run, compared to the empty squares for the NoAGN run. The full distribution, illustrated with a violin plot, is only shown for alternate snapshots for clarity. Dotted horizontal line shows the threshold for the gas being Compton thick ($N_\mathrm{H} = 1/\sigma_\mathrm{T}$). The Reference run is more effective than \fable\ at clearing out gas from the centre of the quasar host galaxy.
    \textit{Fourth row}: Dust mass within 1.5\,kpc, see Section~\ref{Section:Obscuration}, for each run with black holes (solid dark green circles) and for the NoAGN run (empty light green circles), compared to observed values (open black circles) from \citet{Gilli2022} and a measurement from J0100+2802 from \citet{Tripodi2023} (open circle surrounded by black square). The Reference run has a dust mass close to the observed value of J0100+2802, in addition to having a consistent black hole mass and luminosity. 
    \textit{Fifth row}: Distribution of apparent UV magnitudes at 1450\,\AA, along $\sim\!10,000$ sightlines from the black hole, see Section~\ref{Section:Obscuration}. For $z \lesssim 6.3$ the Reference run is brighter than \fable\ due to earlier and more effective feedback.
    \textit{Bottom row}: Probability of observing the quasar along a given sightline with three upcoming wide-field instruments. The quasar in the Reference run has a significantly higher chance of being observed, especially for $z \lesssim 6.3$, due to more effective feedback reducing obscuration compared to \fable. 
}
    \label{fig:BHEddFracfull}
\end{figure*}

\begin{table*}
    \centering
    \caption[Estimate of the black hole luminosity properties.]{A summary of the bolometric luminosity and expected hard X-ray flux (2-10 keV) \citep[calculated from the bolometric correction of][]{Shen2020}, for the primary black hole at a selection of redshifts. Values at each redshift are the median values within $\Delta z < 0.01$, as the accretion rate varies rapidly (see Fig.~\ref{fig:BHEddFracfull}).}    
    \begin{tabular}{c|c|c|c|c}
        
        \hline
    
         & \multicolumn{2}{c}{\fable} & \multicolumn{2}{c}{Reference}\\
        
        $z$ & $L_\mathrm{bol}$ [erg\,s$^{-1}$] & $F_{\mathrm{X}}$ [erg\,s$^{-1}$\,cm$^{-2}$] & $L_\mathrm{bol}$ [erg\,s$^{-1}$] & $F_{\mathrm{X}}$ [erg\,s$^{-1}$\,cm$^{-2}$]\\
        \hline
        6  & $7.0 \times 10^{46}$ & $1.7 \times 10^{-15}$ & $6.9 \times 10^{46}$ & $1.7 \times 10^{-15}$ \\
        7  & $2.1 \times 10^{46}$ & $4.7 \times 10^{-16}$ & $5.1 \times 10^{47}$ & $4.7 \times 10^{-15}$ \\
        8  & $1.1 \times 10^{45}$ & $3.7 \times 10^{-17}$ & $8.6 \times 10^{45}$ & $1.8 \times 10^{-16}$ \\
        9  & $1.8 \times 10^{44}$ & $7.5 \times 10^{-18}$ & $3.0 \times 10^{45}$ & $6.2 \times 10^{-17}$ \\
        10 & $5.7 \times 10^{42}$ & $3.5 \times 10^{-19}$ & $7.8 \times 10^{43}$ & $3.1 \times 10^{-18}$ \\
        11 & $2.4 \times 10^{43}$ & $9.6 \times 10^{-19}$ & $2.0 \times 10^{44}$ & $5.1 \times 10^{-18}$ \\
        
        \hline
    \end{tabular}

    \label{tab:XRayLuminosity}
\end{table*}

In Fig.~\ref{fig:BHEddFracfull} we show the redshift evolution of a number of properties relating to the accretion of the black hole and its observability, with the left panel showing data from the \fable\ run and the right panel showing data from the Reference run.

The top panels in Fig.~\ref{fig:BHEddFracfull} show the Eddington fraction, $f_\mathrm{Edd} = \dot{M}_\mathrm{BH} / \dot{M}_\mathrm{Edd}$, for the largest black hole and its most massive progenitor in both runs. The faded coloured lines show $f_\mathrm{Edd}$ calculated at every timestep the black hole particle is active, with the darker lines on top showing a moving average of 100 timesteps. Observational data points from \citet{KimIm2019} are shown as open circles on the plot, and two data points from \textit{JWST} are shown as gold hexagons \citep{Marshall2023}. As a reminder, \fable\ is capped at $f_\mathrm{Edd} = 1$, and the Reference run is capped at $f_\mathrm{Edd}=2$.  

Above $z=6.5$ the main progenitor in both the \fable\ and Reference runs are accreting near to or above the Eddington rate almost constantly. This demonstrates that in a realistic cosmological environment there is {\it ample} gas supply available in sufficiently rare and massive haloes for black holes to accrete at such large rates. This seems an essential prerequisite to grow large black holes at $z=6$ in such simulations with the seeding prescription we have. There is a significant scatter in the observational data points, and we note that both simulations have a rapid and significant variation in $f_\mathrm{Edd}$ over time, meaning such measurements cannot easily discriminate between different models. Below $z=6.5$, quasar feedback begins to have more of an effect. In both runs, the average accretion rate drops below the Eddington rate, though we note the drops are more pronounced in the Reference run. 

In the second row of Fig.~\ref{fig:BHEddFracfull}, we show the bolometric luminosity of the black holes. The bolometric luminosity is calculated using 
Equation~(\ref{Eqn:Lbol}), so we note that the Eddington rate cap is also present in these panels. In the \fable\ run we see that while there is a significant variation over time, the luminosity gradually increases as the black hole grows, with the average reaching a peak of $10^{47}$\,erg\,s$^{-1}$ by the end of the simulation. \fable\ is consistent with many quasars with moderate luminosities, however the produced black hole luminosities are dramatically below a number of observed high-redshift quasars.

Switching to the Reference run we see a different picture, as the earlier growth of the black hole and increased Eddington limit lead to significantly higher luminosities than in \fable. The average luminosity is consistent with the brightest quasars at higher redshifts, and shows how a $10^9\,\mathrm{M}_\odot$ black hole at $z\sim7.5$ can grow to be $10^{10}\,\mathrm{M}_\odot$ by $z=6$. We note that in the entire redshift range plotted for the Reference run, the luminosity rarely drops below $10^{46}$\,erg\,s$^{-1}$, is on average above $10^{47}$\,erg\,s$^{-1}$ and actually spends a significant fraction of its life at $10^{48}$\,erg\,s$^{-1}$. 

No other quasars with $L_\mathrm{bol} \geq 10^{48}$\,erg\,s$^{-1}$ have yet been observed, which could be due to several reasons. Firstly, large enough haloes are very rare at high-redshift, so a larger survey volume may be needed - the Vera Rubin Observatory, \textit{Roman}, and \textit{Euclid} may probe the existence of such rare and bright objects. Moreover, the bright quasars we see may be the tip of the proverbial iceberg of high-redshift supermassive black holes. With future surveys that can observe quasars pushing both to $z \gtrsim 7.5$ and lower luminosities, $L_\mathrm{bol} \lesssim 10^{46}$\,erg\,s$^{-1}$ -- a revolution already underway with \textit{JWST} -- we will be able to gain unique constraints both on the rarity of high-redshift quasars (i.e. their halo occupation fraction), and on the assembly history of these ultramassive black holes, where it may be possible to more robustly disentangle growth contributions from gas accretion and black hole mergers. Furthermore, it is worth mentioning that the quasar variability may be larger than found in our simulations and that non-isotropic quasar emission may further complicate this picture, with very bright quasars then preferentially observed `down-the-barrel'.  

Additionally, the growth of such a massive black hole might be obscured, despite the incredibly high bolometric luminosity. In the third row of Fig.~\ref{fig:BHEddFracfull} we show the evolution of the distribution of $N_\mathrm{H}$ column densities at a distance of 5\,kpc for $\sim\!10,000$ sightlines around the black hole (see Section~\ref{Section:Obscuration}). Solid squares and violin plots show the median and distribution of column densities for the \fable\ (left) and Reference (right) runs. In both left- and right-hand panels, open squares show the median column density for the NoAGN run. The first thing to note is that the NoAGN run has significantly higher median column densities than the \fable\ run below $z\sim7$, and higher than the Reference run for almost all redshifts shown. In fact, the NoAGN run straddles the horizontal dotted line showing a Compton thick column density throughout the simulation. The lower $N_\mathrm{H}$ columns in the runs with black holes are a clear indicator of AGN feedback in these models, with the breadth of the distribution also widening as gas is ejected from the central galaxy. In the Reference run, the median column density fairly consistently decreases over time from $z\sim7$, as repeated feedback episodes have a cumulative effect.

The qualitative trends we see in $N_\mathrm{H}$ column density evolution are similar to that seen by \citet{Vito2022}, who also find a broader distribution of column densities and lower median when their AGN feedback kicks in, particularly in runs utilising biconical outflows. Interestingly, we note that in both \fable\ and the Reference run, feedback energy is injected isotropically, which \citet{Vito2022} find hinders their black hole growth and diminishes the impact of feedback. This implies that more anisotropic injection of feedback energy could help black holes grow faster in the early Universe. \citet{Ni2020} also find that AGN feedback broadens the distribution of column densities around quasars in \textsc{BlueTides}. In some cases this can lead to significantly lower column densities than in our simulations, though we note this could be because they are looking at smaller haloes than ours, which have shallower gravitational potential wells.

The fourth row of Fig.~\ref{fig:BHEddFracfull} shows the estimated total dust mass within 1.5\,kpc, assuming a 15 per cent dust-to-metal fraction \citep[like in][]{DiMascia2021,Vito2022}, with the runs with black holes shown as solid circles and the NoAGN run shown as empty circles. The 1.5\,kpc radius was chosen to match the largest measured dust radius in the sample of \citet{Gilli2022}, and we compare our results to their measurements (open, black circles). The dust masses in the runs with and without black holes are similar at the highest redshifts considered, with differences appearing when feedback kicks in after $z=7$ in the \fable\ run, and after $z=7.5$ in the Reference run. The dust masses then decrease by nearly an order of magnitude by $z=6$, while the mass in the NoAGN decreases slowly by a much smaller amount. This corresponds well to the decrease in hydrogen column density seen in the panel above, as metals are ejected from the host galaxy. Dust will additionally be destroyed by feedback episodes, however we only follow this in a simple manner, by considering only the `cold' metal mass as described in Section~\ref{Section:Obscuration}. Compared to the observational data, we find both the NoAGN and \fable\ runs can match the amount of dust in a number of observed quasar host galaxies. 

In contrast, the Reference run - which has a larger black hole mass than most of the sample of \citet{Gilli2022} - has considerably lower dust masses than the NoAGN and \fable\ runs. In fact, the observed data point surrounded by a square shows a measurements of the dust mass of the host galaxy of J0100+2802, the brightest high-redshift quasar. This recent measurement used observations in Band 9 of ALMA to estimate $T_\mathrm{dust}=48.4\pm2.3$\,K and the emissivity index $\beta=2.63\pm0.23$, leading to an inferred dust mass of $2.29\pm0.83\times10^7$\,M$_\odot$ \citep{Tripodi2023}. Both our host galaxy's SFR (see Section~\ref{Results:SFR}) and dust mass at $z=6.3$ in the Reference run reasonably match the observational estimates from \citet{Tripodi2023}, though we note as their dust continuum emission is unresolved there is uncertainty on which spatial scale to consider the dust mass in the simulation. 

While we have successfully grown a $10^{10}$\,M$\odot$ black hole in our simulated halo, it could well have had a significantly different evolution history to the host of J0100+2802. The aim of this paper was not to produce an exact analogue of J0100+2802, so it is interesting we can approximately reproduce the observed properties of the host (albeit with caveats on our dust mass estimates, discussed in Section~\ref{Section:Obscuration}). In Fig.~\ref{fig:BHEddFracfull}, we can see that AGN feedback can reduce the estimated dust mass in the quasar host, in both the \fable\ and Reference runs, rapidly. In particular, in the Reference run (right panel), between $z=6.1$ and the end of the simulation at $z=5.75$, the dust mass drops by an order of magnitude in considerably less than 100\,Myr -- black hole feedback can rapidly eject and destroy dust. 

\begin{figure*}
    \centering
    \includegraphics[width=\linewidth]{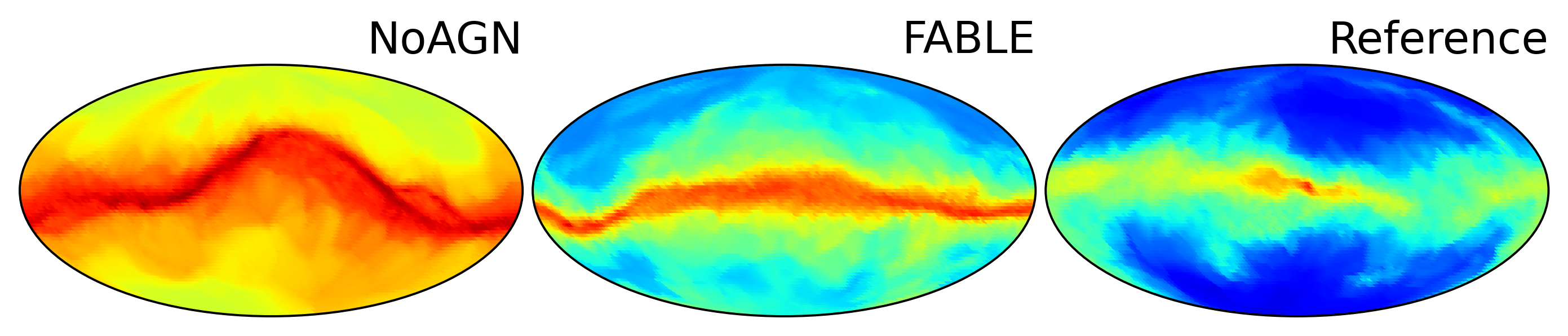}
    \includegraphics[width=\linewidth]{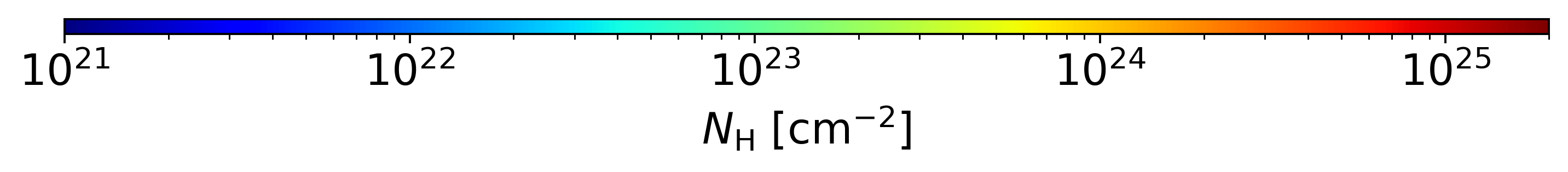}
    \includegraphics[width=\linewidth]{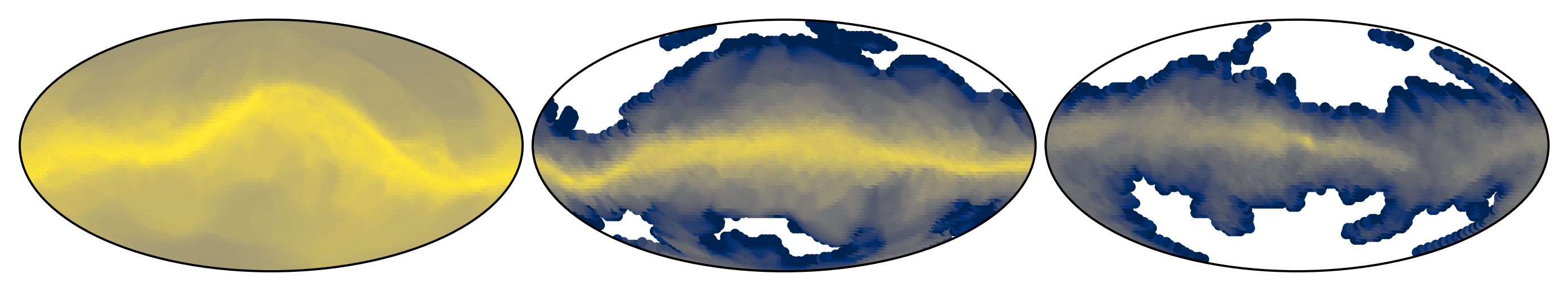}
    \includegraphics[width=\linewidth]{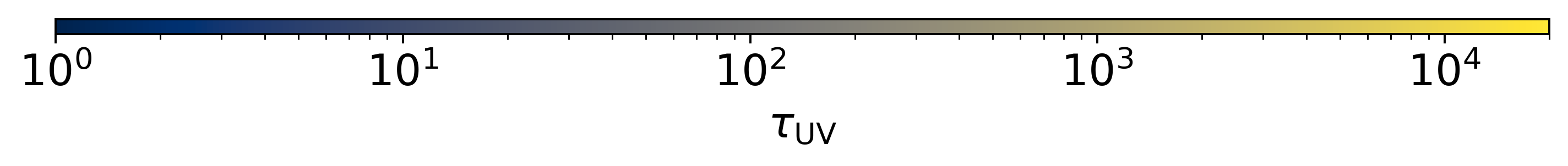}
    \caption{Mollweide projection maps of the central 5\,kpc of the central galaxy at $z=6.1$ in the NoAGN (left), \fable\ (centre) and Reference (right) runs. We show the hydrogen column density (top) and UV optical depth $\tau_\mathrm{UV}$ (bottom), both of which are calculated along the $\sim\!10,000$ rays described in Section~\ref{Section:Obscuration}. The increase in the strength of feedback can clearly be seen going from left to right, with more low-density channels in the surrounding gas.}
    \label{fig:mollweide}
\end{figure*}

In the fifth row of Fig.~\ref{fig:BHEddFracfull} we show the distribution of UV apparent magnitudes (rest-frame 1450\AA) along the same sightlines as the column densities (see Section~\ref{Section:Obscuration} for full details of this calculation). We note that these distributions are very broad and extend far beyond what is, or will be, feasible with observations. The median magnitude of the quasar varies significantly over time as the halo undergoes mergers and feedback events. Throughout, though, the median in the Reference run tends to be below that of the \fable\ run, due to earlier, more powerful feedback. The impact of feedback in the Reference is particularly noticeable, after $z\sim6.3$, where the median magnitude becomes consistently lower due to the clearing effect of the quasar. We note that this shows obscuration in the UV, corresponding to high UV optical depths (see Fig.~\ref{fig:mollweide}). While the dust opacity in the infrared is $\sim\!100$ times smaller than in the UV, this will still lead to a large number of sightlines with infrared optical depths $\tau_\mathrm{IR} \gg 1$, suggesting infrared radiation pressure on dust is likely to be an important mechanism of feedback \citep[see][for an investigation of this process]{Costa2018}. 

We further note that when observing objects at such redshifts, the bandpass of X-ray observatories such as \textit{AXIS} or \textit{Lynx} will allow for the detection of harder X-ray photons, which are considerably less impacted by obscuration (though it is worth pointing out that these instruments are not primarily designed to survey large areas of sky). In Table~\ref{tab:XRayLuminosity} we show the bolometric luminosity and hard X-ray flux \citep[using the correction of][]{Shen2020} of the largest black hole in the simulation at redshifts 6-11. These values are the median value during a time $\Delta z < 0.01$ of each redshift. We can compare these values to the expected hard-band sensitivities of a number of proposed future surveys, from \citet{Habouzit2022a} and using their K-correction. The most sensitive proposed survey is the \textit{Lynx} DEEP survey, which is forecast to spend 4\,Ms looking at a small patch (0.11\,deg$^2$) of sky. This would have a flux limit of $\sim4.2\times^10^{-17}$\,erg\,s${-1}$\,cm$^{-2}$ and so would be able to detect the quasar itself in both runs at any of the given redshifts, assuming we viewed the quasar down a Compton-thin sightline. In comparison, the proposed shallower WIDE surveys of \textit{ATHENA} and \textit{AXIS} would survey $\sim50$\,deg$^2$ of sky for $\mathcal{O} 10$\,ks, to a depth of $\sim 1\times10^{-16}$\,erg\,s${-1}$\,cm$^{-2}$. This implies the \fable\ run could only be detected at $z\lesssim7$, whereas the Reference run would be visible for longer, $z\lesssim8$. See Section~\ref{Section:XRayResults} for further discussion of X-ray observations of quasar hosts.

This change in apparent magnitudes can be translated into a probability of being observed - taking the number of sightlines with a magnitude exceeding the threshold value of a particular survey divided by the total number of sightlines. Note that this is a chance of one object being observed at different orientations, not a statement on the observability of a population of quasars. A rest-frame wavelength of 1450\AA\ at $z=6$ translates into a y-band observation of the Vera Rubin Observatory \citep[VRO, ][]{LSST} and \textit{Euclid} \citep{EuclidHighzQuasars,Euclid}, or the F106 filter of the \textit{Nancy Grace Roman Space Telescope} \citep{ROMAN}. Taking the limiting magnitudes for point sources with these telescopes, we calculate the probability that the simulated object would be observed at each redshift. At higher redshifts the filter that the rest-frame UV would fall into will change, for example the VRO will lose sensitivity to a rest-frame wavelength of 1450\AA\ at around $z\sim6.24$. However we note that the limiting magnitudes for the bands of \textit{Euclid} and \textit{Roman}, within the redshift range on this plot, are similar. 

We show the calculated probabilities of observing the quasars in the bottom panel of Fig.~\ref{fig:BHEddFracfull} for the full 10-year Legacy Survey of Space and Time (LSST) with the VRO (limiting y-band magnitude of 24.9), the \textit{Euclid} Deep survey (1$\sigma$ limiting y-band magnitude of 26.0), and with one hour of integration with \textit{Roman} (limiting magnitude of F106 filter in 1 hour is 28.1). The \fable\ quasar remains very obscured until after $z\sim7$, after which enough metals are cleared by feedback to allow the quasar to become visible along some sightlines. After this, the probabilities vary over time as inflows and outflows compete, but generally hover around 30-40 per cent. The broad distribution of UV magnitudes and probability of observing the quasar is also in qualitative agreement with \citet{Vito2022}, who find that only the 50 per cent of their sightlines that were least extincted showed magnitudes consistent with high-redshift quasar observations. 

For the Reference run there is a considerably higher chance of observing the quasar throughout, due to earlier growth and feedback. In the Reference run, the quasar becomes less obscured over time due to the cumulative impact of feedback, with a probability of $\sim\!50$ per cent by $z=6$. In the wake of mergers (e.g. around $z\sim7.1$) the difference between the runs narrows and both become more obscured. We note that evidence from recent \textit{JWST} observations and simulations with on-the-fly radiative transfer suggest bursty star formation after a merger can lead to the clearing of neutral gas that can allow Ly$\alpha$ photons to escape \citep{Witten2023}.

The shape of the evolution of probabilities is largely similar between the different surveys shown (at least for when a rest-frame wavelength of 1450\AA\ is detectable by the VRO), however it is important to mention that the \textit{Roman} limiting magnitude used here is for only 1 hour of observations. With deep, co-added stacks, the magnitude could be pushed further, to $\sim\!29$ mag \citep[e.g.][]{Rose2021}.

It is also worth noting that we performed the same calculation with the NoAGN run, even assuming a constant bolometric luminosity of $10^{48}$\,erg\,s$^{-1}$ at every redshift, and the probability of observation with any survey remained zero. This strongly implies that feedback from the quasar is \textit{required} for us to observe such quasars at all in the rest-frame UV \citep[which has been found with different codes and physical models, e.g.][]{Costa2018,Costa2022}. These results also tally well with short inferred quasar lifetimes from observations of proximity zones, which require significant amounts of obscured accretion at high redshift \citep[e.g.][]{Eilers2017,Satyavolu2023}.

Our results about the obscuration of the central quasar are additionally illustrated in Fig.~\ref{fig:mollweide}, which shows maps of gas properties at $z=6.1$ along the same rays we previously used. These maps have been rotated to approximately line up with the angular momentum vector of the central 2\,kpc of the galaxy, though we note the different runs do not exactly line up with each other due to different gas kinematics. 

The top row of panels shows hydrogen column density maps. While all three runs have a central disc structure of gas around the galaxy centre, the densities of that disc are drastically different in each run. The NoAGN run has very high column densities, almost always in excess of $10^{24}$\,cm$^{-2}$, surrounding the halo centre. The feedback of the \fable\ run is sufficient to expel much of this gas, with only the galaxy disc retaining such high column densities. Here, column densities along the minor axis of the galaxy are reduced to $\sim\!10^{22}$\,cm$^{-2}$. In the Reference run this trend continues, with feedback ejecting significant amounts of gas along the minor axis of the galaxy. This reduces column densities in all directions, with the exception of the very centre of the disc, by approximately an order of magnitude over the \fable\ run. 

In the second row of Fig.~\ref{fig:mollweide} we can see how a reduction in column density leads to a change in the probability of observing the quasar as we show maps of UV optical depth, $\tau_\mathrm{UV}$. The optical depth of the NoAGN run is >100 along all sightlines and so remains completely obscured in the rest-frame UV. The \fable\ and Reference runs still have many obscured lines of sight through the galactic disc, but with increasing numbers of `holes' (where $\tau_\mathrm{UV} \to 0$) as AGN feedback becomes more powerful. We note that many sightlines have very high UV optical depths, $\tau_\mathrm{UV}>10^4$, implying such a quasar could even be obscured in the infrared too.

\subsection{Impact on star formation} \label{Results:SFR}

\begin{figure}
    \centering
    \includegraphics[width=\linewidth]{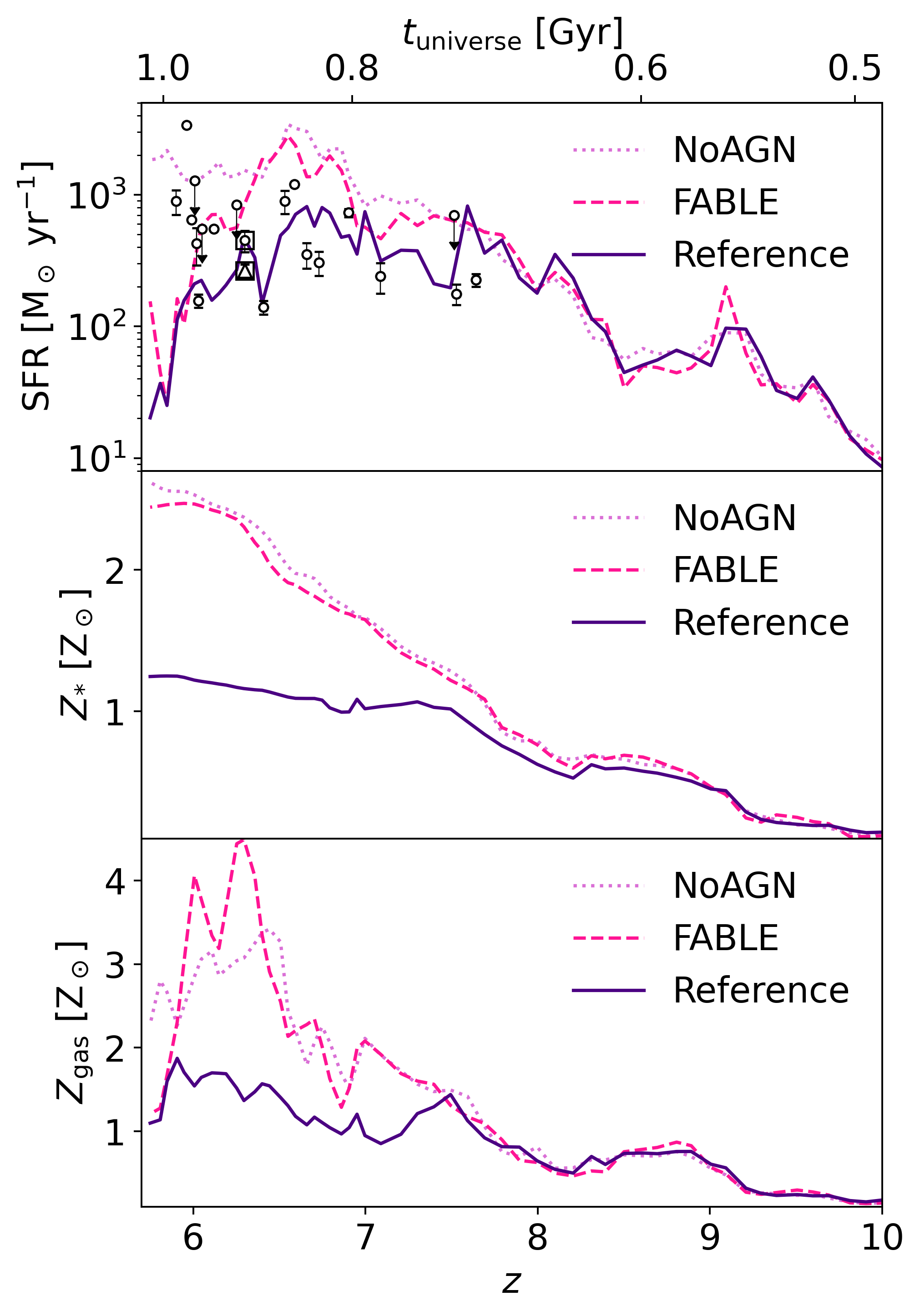}
    \caption[Star formation rate and stellar-to-halo mass ratio of the supermassive black hole host galaxy over time.]{\textit{Top}: Star formation rate as a function of redshift for all simulations. Note that the AGN feedback in the fiducial \fable\ model has a negligible effect on star formation until much later than the Reference run. Observational data points are shown as open circles, from the sources cited in Section~\ref{Section:BHMass}, with errors where available. As a reminder, most of these black holes have inferred masses in excess of $2\times10^9$\,M$_\odot$, with the exception of a few sources at $z>7$.  The black squares highlight data from J0100+2802, the largest observed high-redshift quasar, with the higher SFR coming from \citet{KimIm2019} and the lower SFR coming from ALMA dust measurements \citep{Tripodi2023}. The Reference run has a SFR consistent with the accurate ALMA-based SFR of J0100+2802.
    \textit{Middle}: Evolution of the stellar metallicity of the quasar host galaxy, calculated as the mass-weighted average metallicity of all stars within twice the stellar half-mass radius. The Reference run has considerably lower stellar metallicity in the central galaxy, due to both the lower stellar mass and expulsion of metal-rich gas.
    \textit{Bottom}: Evolution of the galaxy gas metallicity, again calculated as the mass-weighted average metallicity of all gas cells within twice the stellar half-mass radius. Similarly to the middle row, the Reference run has a lower average gas metallicity than the NoAGN or \fable\ runs from $z=7.5$ onwards.}
    \label{fig:Mstargrowth}
\end{figure}

\begin{figure}
    \centering
    \includegraphics[width=0.95\linewidth]{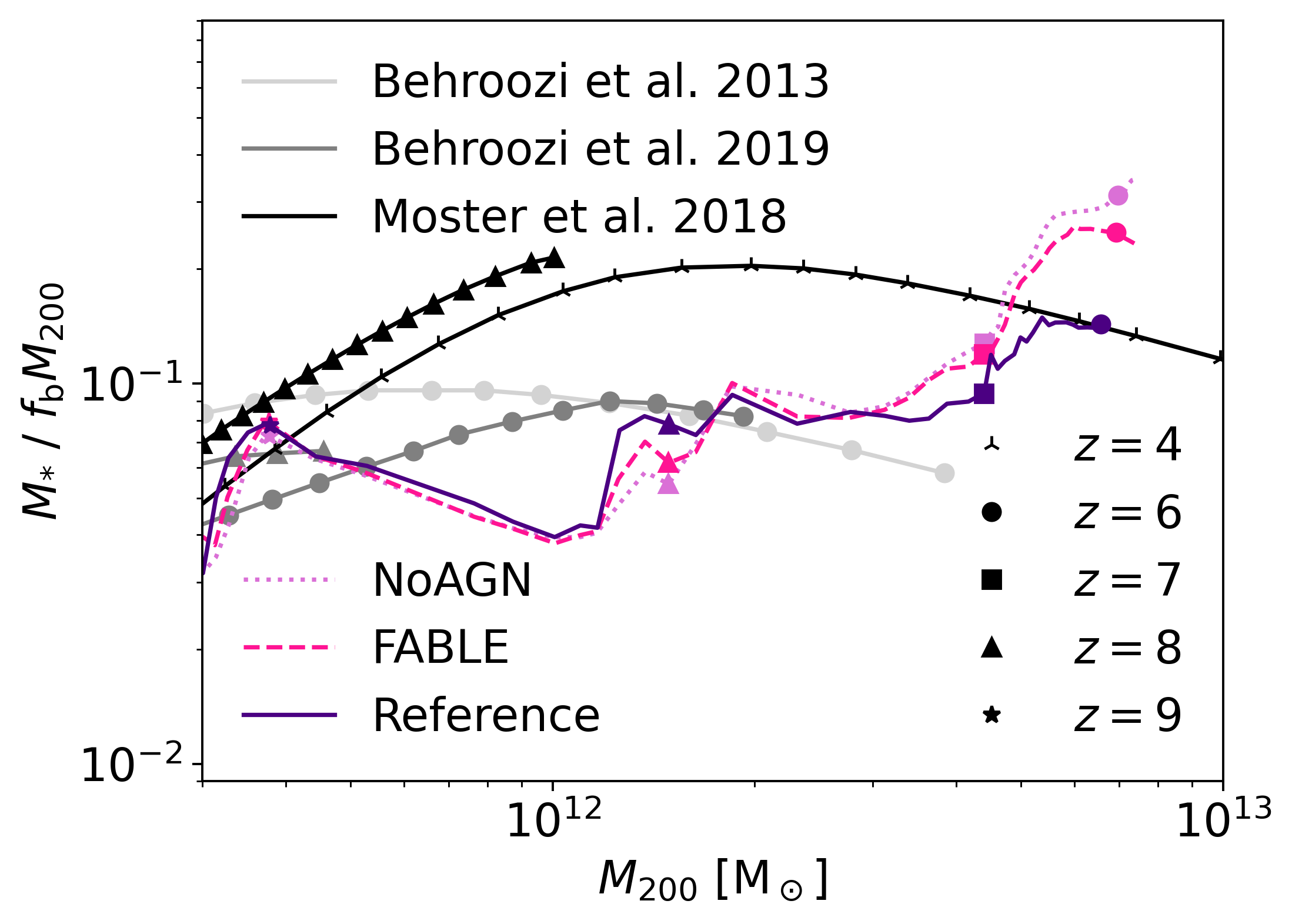}
    \caption{Ratios of stellar mass to (expected) baryonic mass as a function of halo mass for the three runs in this work, showing their evolution in this parameter space. The values for each run at $z=9,8,7$ and $6$ are shown by the star, triangle, square and circle symbols, respectively. We also show the median ratio of this quantity from a number of models (though we note these all assume \textit{Planck} cosmology) from \citet{Behroozi2013,Behroozi2019,Moster2018}. The Reference run is clearly more effective at regulating star formation compared to \fable\ or a run without black holes.}
    \label{fig:MstarMhalo}
\end{figure}

The feedback required to clear the surroundings of such a massive and bright AGN could potentially have a significant impact on the growth of the host galaxy itself. The top panel of Fig.~\ref{fig:Mstargrowth} shows the star formation rate (SFR) of the central galaxy in the largest halo as a function of redshift. Observational data points from \citet{KimIm2019} are shown as open circles. We also show a recent measurement of the SFR of J0100+2802 using ALMA dust measurements with a triangle \citep{Tripodi2023}.

The three runs are largely consistent until just after $z=8$, when feedback in the Reference run causes a drop in SFR. The \fable\ run continues to track the NoAGN run until around $z=6.5$, when its black hole grows large enough to affect star formation too. At $z=6$, the SFR of both runs with black holes is two orders of magnitude below the NoAGN run. The observed SFR values typically sit below the NoAGN run, suggesting some amount of AGN feedback is needed to match them. An important thing to note is that when the AGN kicks in, in both the \fable\ and Reference runs, the SFR can be lowered by an order of magnitude within a short timescale (<100\,Myr). This is similar to the dramatic drop in estimated dust mass seen in Fig.~\ref{fig:BHEddFracfull}. A wide range of observed SFRs could therefore be observed, depending on the growth history of the halo and its primary black hole. The SFR estimates for the host of J0100+2802 (data points surrounded by squares in Fig.~\ref{fig:Mstargrowth}), from both \citet{KimIm2019} and \citet{Tripodi2023}, are in reasonable agreement with the Reference run, similarly to the estimated dust mass.

In the middle panel of Fig.~\ref{fig:Mstargrowth}, we show how this change in the feedback and star formation history can affect the properties of the stellar component by looking at the mass-weighted average metallicity of the central galaxy (within two times the stellar half-mass radius). We can see that despite the \fable\ halo hosting a $10^9$\,M$_\odot$ black hole by $z=6$, there is little change in the average stellar metallicity of the central galaxy. In contrast, the Reference run shows a factor of two decrease in the average stellar metallicity in the galaxy by $z=6$. This difference seems to appear after about $z=9$, when the Reference run black hole reaches around $10^8$\,M$_\odot$. This effect arises due to the lower stellar mass, and hence lower metal production, of the host galaxy by $z=6$ (see Table~\ref{tab:FinalValues}). Looking at the bottom panel of Fig.~\ref{fig:Mstargrowth} we can see additional signs of this, with lower gas-phase metallicity in the Reference run from $z=7.5$ onward, due to both lower initial production of metals and their subsequent ejection from the galaxy. Measurements of the gas and stellar metallicity in high-redshift quasar hosts, for example with \textit{JWST}, could therefore be an interesting and useful test for such models of feedback.

We look in more detail at how these changes in stellar mass affect the stellar-to-halo mass relation in Fig.~\ref{fig:MstarMhalo}. We show the evolution of this quantity as a function of halo mass for all of our runs, with star, triangle, square and circle symbols indicating $z=9, 8, 7$ and $6$, respectively. For reference, we also show the median ratio of stellar mass to baryonic mass from \citet{Behroozi2013} at $z=6$, from \citet{Behroozi2019} at $z=6$ and $z=8$, and from \citet{Moster2018} at $z=4$ and $z=8$. We note all of these models assume \textit{Planck} cosmology rather then the older \textit{WMAP} cosmology assumed for our simulated halo. In addition, the very massive halo we are considering in this work is higher than the upper end of these functions, so this comparison should be taken with caution. 

Nevertheless, the $z=6$ stellar masses of the NoAGN and \fable\ runs are clearly well above the expected value of this ratio. This implies that not only do we require AGN feedback to reduce the rate of star formation early on in the Universe for very massive haloes, but also that the fiducial \fable\ model is insufficient to do so. This is despite \fable\ better matching (or even overquenching) the SFR of galaxy clusters at low redshift, further suggesting the redshift evolution of AGN feedback is incorrect in \fable\ (and in the original Illustris, on which \fable\ is based). At low redshift, there is also a good match in stellar mass fraction between \fable\ and observed haloes \citep[i.e.][]{FABLE1,Kravtsov2018}, the implication therefore being that the evolution of the stellar mass fraction is also not correct in \fable. 

At $z=8$, the simulated halo would likely be above the extrapolated stellar-to-baryonic mass fraction of \citet{Behroozi2019}, but below that of \citet{Moster2018}, highlighting the significant uncertainties in such models. We note that future accurate observational estimates of the stellar masses of such bright quasars host will provide important constraints on theoretical models of AGN feedback. It is worth noting that the NoAGN and \fable\ runs at $z=6$ lie above even the $z=4$ ratio from \citet{Moster2018}, suggesting more significant feedback would be needed to bring them back onto such a relation.

By $z=6$, the stellar mass fraction of the Reference run is at $\sim\!15$ per cent, much lower than the $\sim\!25-30$ per cent in the NoAGN/\fable\ runs. This is closer to the (extrapolated) median $z=6$ fractions from \citet{Behroozi2013,Behroozi2019}, though we note such a large overdensity as the one we follow may already lie off of such a relation, regardless of feedback. 

\begin{figure}
    \centering
    \includegraphics[width=\linewidth]{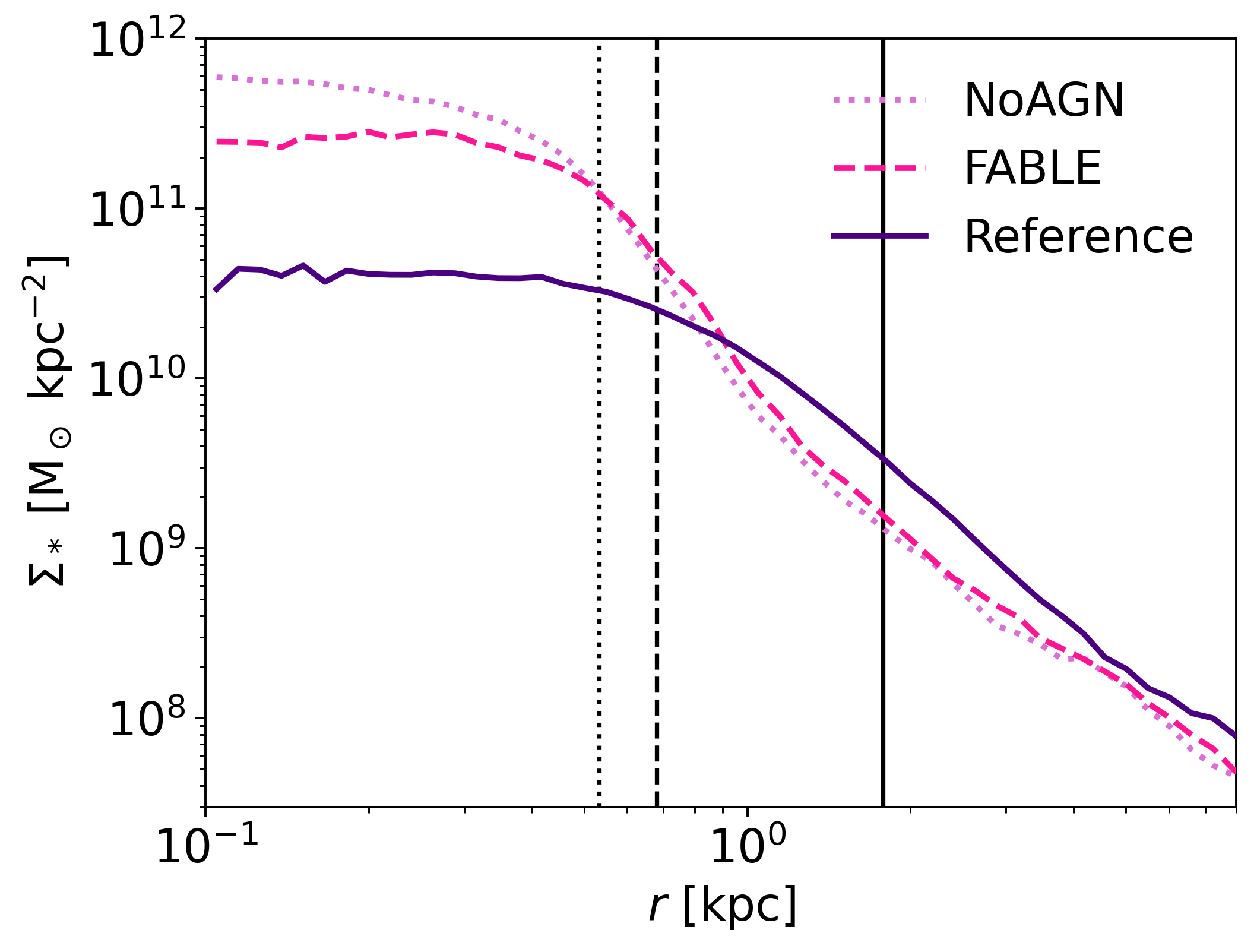}
    \caption{Stellar surface density radial profile of the quasar host galaxy at $z=6$, in the NoAGN (dotted), \fable\ (dashed), and Reference (solid) runs. Vertical lines using the same line styles show the stellar half-mass radius for each of the three runs (see Table~\ref{tab:FinalValues}). The distribution of stars changes significantly with feedback, with an increased black hole mass correlating with a lower central density and broader stellar distribution.}
    \label{fig:StellarDensity}
\end{figure}

\begin{figure*}
    \centering
    \includegraphics[width=\linewidth]{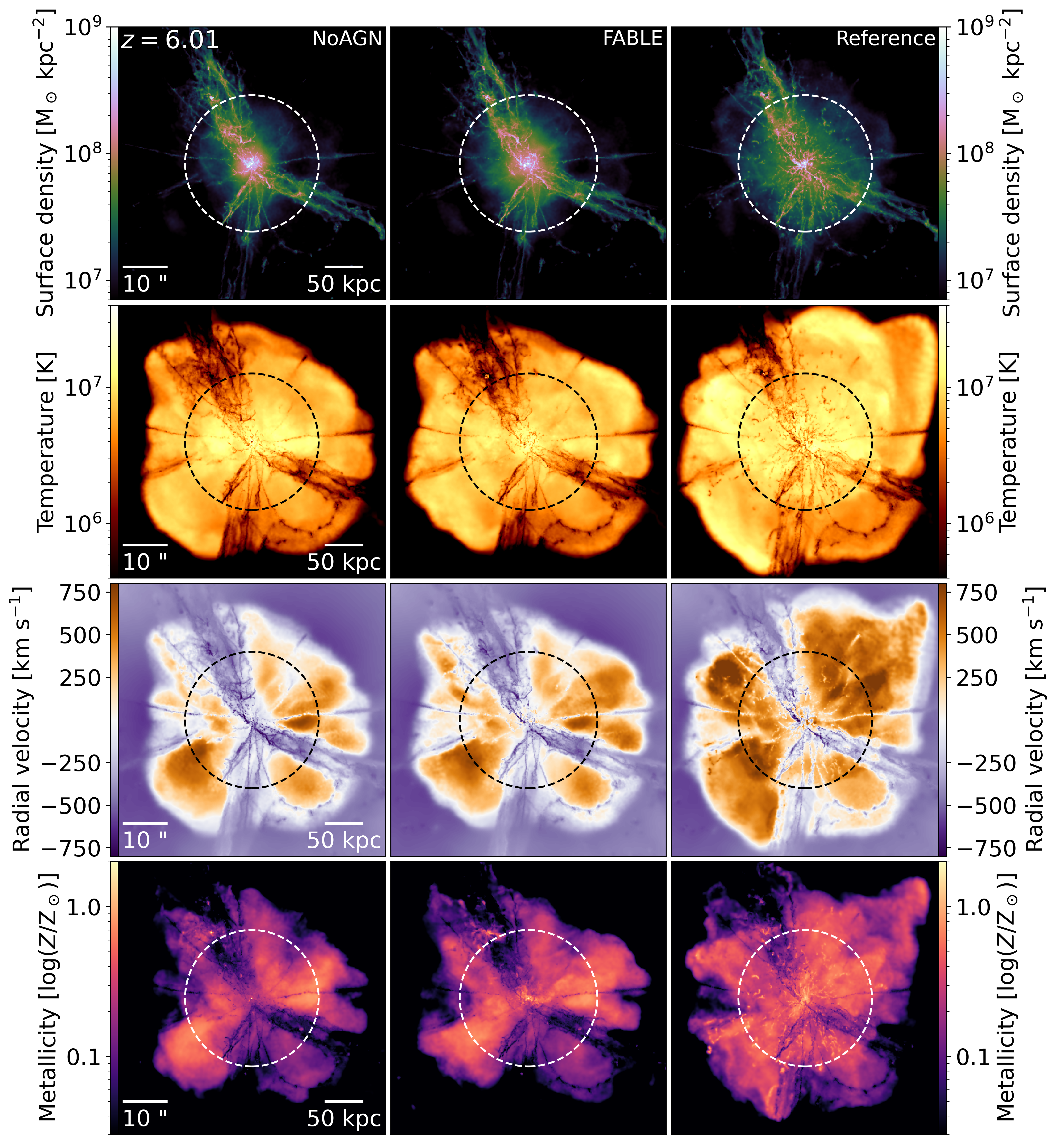}
    \caption[Maps of the gaseous halo around the quasar at $z=6$.]{Maps of the gaseous halo around the quasar at $z=6$ showing, from top to bottom, surface density, and mass-weighted average temperatures, radial velocities, and metallicities. From left to right the columns show the NoAGN, \fable, and Reference runs. Dashed circles show the virial radius in each panel. The AGN in the \fable\ run has little impact on the CGM by $z=6$, whereas the feedback in the Reference run expels significant amounts of gas at high velocities deep into the CGM.}
    \label{fig:z6maps}
\end{figure*}

\subsection{Galaxy structure} \label{Section:StellarStructure}

While we have already discussed how the total stellar mass is reduced when black holes can grow and provide feedback earlier, we also find that the different models of AGN feedback can also lead to a different stellar structure of the host galaxy. We investigate this in Fig.~\ref{fig:StellarDensity}, in which we show radial profiles of the stellar surface density for the central galaxy at $z=6$ for the NoAGN (dotted), \fable\ (dashed), and Reference (solid) runs. The NoAGN has the most centrally concentrated stellar density profile, reaching a peak of $\sim\!10^{12}$\,M$_\odot$\,kpc$^{-2}$ at a scale of 0.1\,kpc. In the \fable\ run, which does affect star formation significantly but only at later times closer to $z=6$, we see the central stellar density profile is flatter, reflecting the reduction in star formation on sub-kpc scales. Outside of $\sim\!0.6$\,kpc, the shape of the stellar density profiles in the NoAGN and \fable\ runs are very similar. 

Looking at the stellar density profile of the Reference run in Fig.~\ref{fig:StellarDensity}, we see some notable differences. While the profile is flat in the galaxy centre, similarly to the \fable\ model, the central density is decreased further due to powerful AGN-driven winds driving star forming gas out of the innermost regions. More interestingly, on larger scales ($r > 1$\,kpc), we find an \textit{enhancement} in stellar densities relative to the NoAGN and \fable\ runs. The strong, early black hole feedback present in the Reference run seems to broaden the stellar distribution of the halo, as well as lower central densities, which could both be interesting probes using future observations of quasar hosts from \textit{JWST}. A similar effect occurs in the simulations of \citet{vanderVlugt2019}, which is shown to be caused by fluctuations in the gravitational potential caused by bursty AGN feedback. 

\subsection{Impact on the gaseous halo}

We visually inspect the impact of AGN feedback on the CGM properties of our different runs at $z=6$ in this section by presenting maps of a number of quantities.

\subsubsection{Basic properties}

\begin{figure*}
    \centering
    \includegraphics[width=\linewidth]{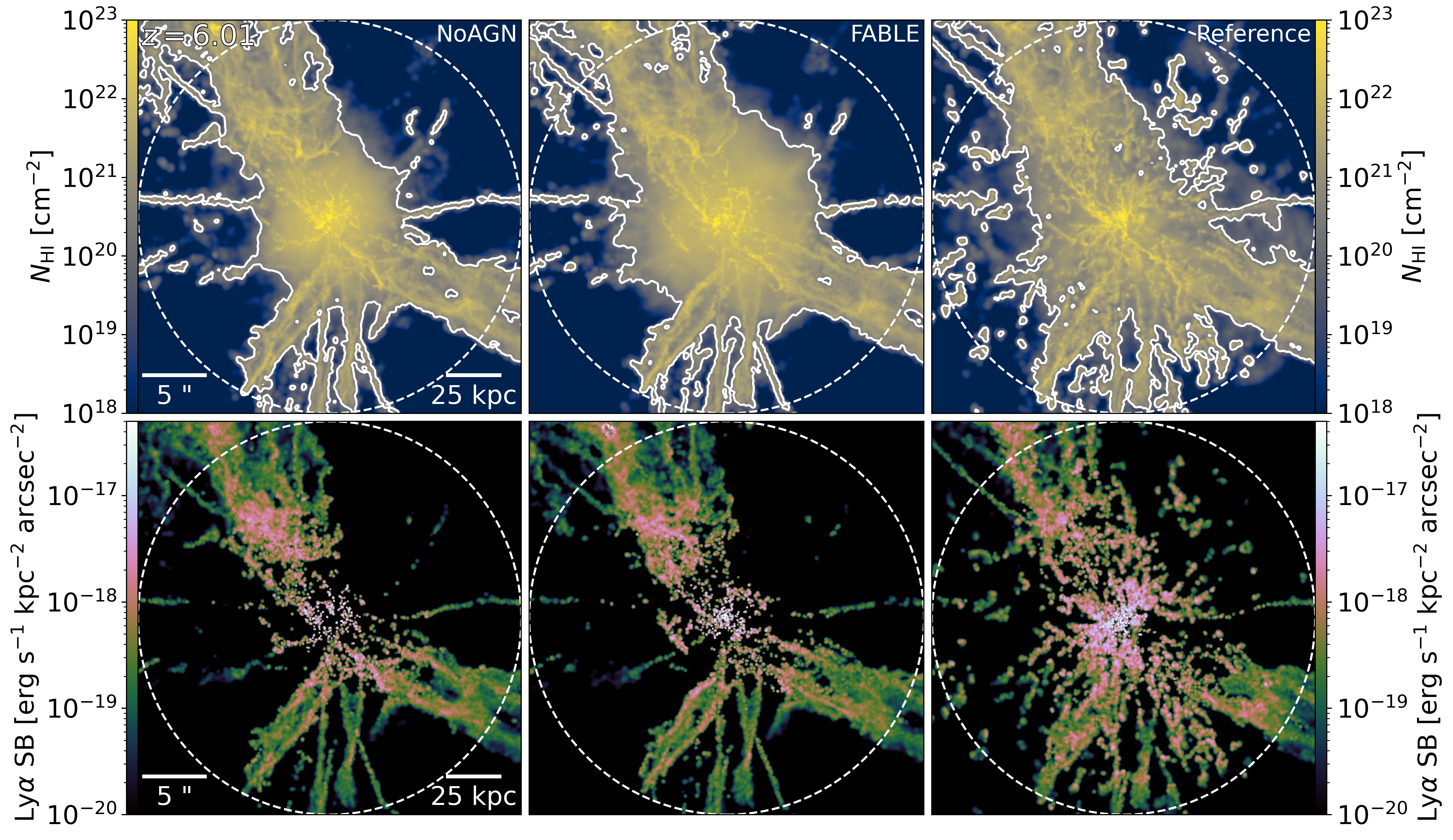}
    \caption[Maps of H\textsc{i} and collisional Ly$\alpha$ in the gaseous halo around the quasar at $z=6$.]{Maps of H\textsc{i} column densities (top) and collisional Ly$\alpha$ surface brightness (bottom) around the quasar at $z=6$. From left to right the columns show the NoAGN, \fable, and Reference runs. Dashed circles show the virial radius in each panel. In the top panel, white contours show the boundary of $10^{20.3}$\,cm$^{-2}$, the threshold for a Damped Ly$\alpha$ system (DLA). The total Ly$\alpha$ luminosity (from collisional ionisation only) for each run is $7.64\times10^{43}$, $7.90\times10^{43}$, and $1.19\times10^{44}$\,erg\,s$^{-1}$ for the NoAGN, \fable\ and Reference runs, respectively. The small impact of the \fable\ feedback is more visible than in Fig.~\ref{fig:z6maps}, with a broader H\textsc{i} distribution, though the AGN feedback in the Reference run is more powerful and so ejects gas clumps to large distances from the central galaxy.}
    \label{fig:z6linemaps}
\end{figure*}

\begin{figure*}
    \centering
    \includegraphics[width=\linewidth]{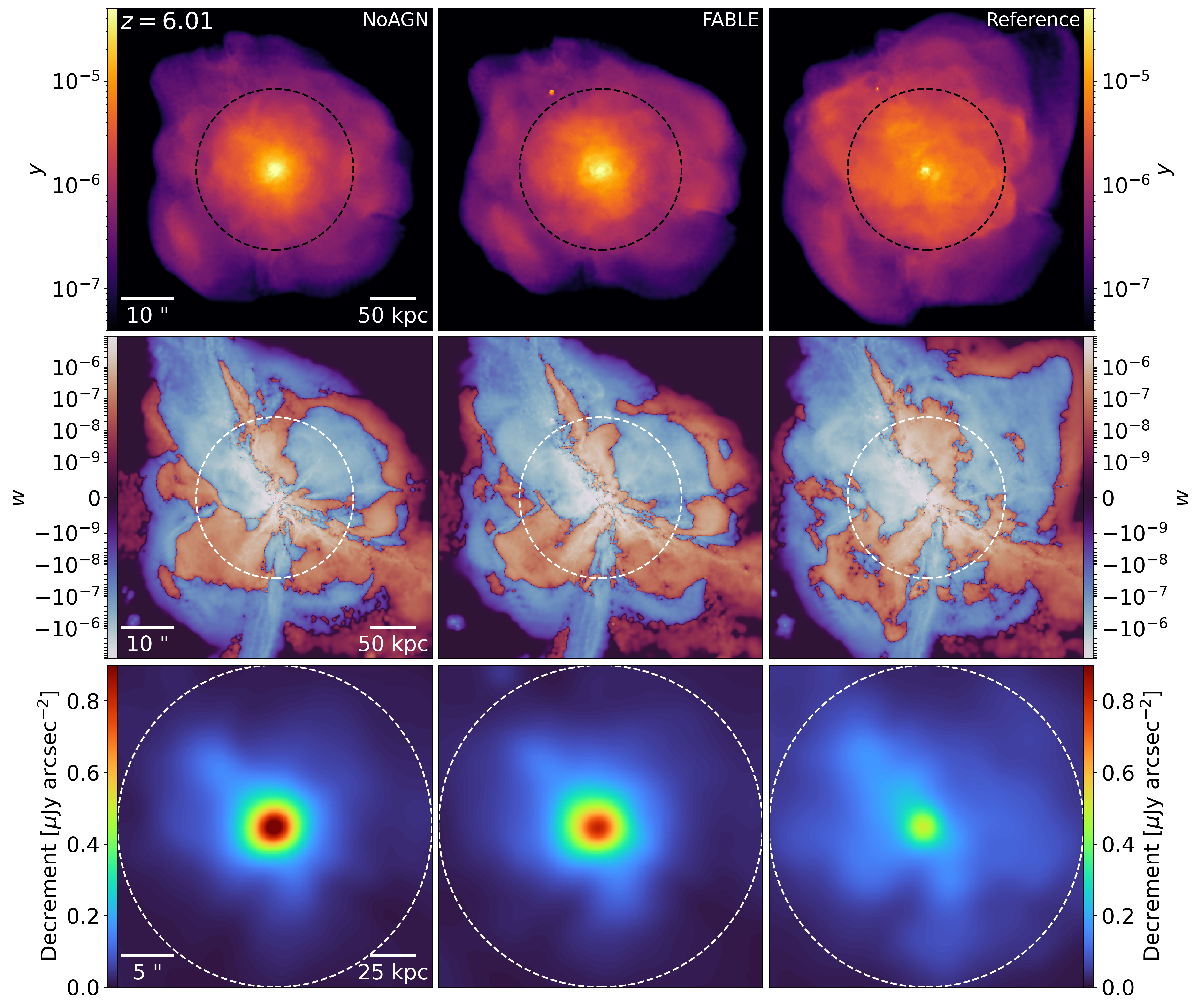}
    \caption[SZ maps at $z=6$ of the gaseous halo hosting the largest supermassive black hole in our simulations.]{SZ maps at $z=6$ of, from top to bottom, Compton-$y$ parameter from the thermal SZ effect, $w$ parameter from the kinetic SZ effect, and the combined SZ decrement. The latter has been smoothed on a scale of 1 arcsecond, to mimic the beam of an instrument like ALMA. From left to right the columns show the NoAGN, \fable, and Reference runs. Dashed circles show the virial radius in each panel. The redistribution of central gas in the Reference run is shown very clearly in the tSZ and combined SZ signals, with obviously lower central pressures and a broader signal.}
    \label{fig:z6SZmaps}
\end{figure*}

Firstly, in Fig.~\ref{fig:z6maps} we show, from top to bottom, gas surface density, and mass-weighted average temperature, radial velocity and metallicity. It is evident from the left-hand and central columns that the NoAGN and \fable\ runs are very similar in the metrics we have plotted, again pointing to the ineffectiveness of the fiducial \fable\ model to impact this halo at such high redshift. In the very centre of the halo there are some initial signs of higher metallicities and outflow velocities as the \fable\ feedback has kicked in recently, and of a slightly broader gas distribution (seen more clearly in the zoomed in H\textsc{i} maps in Fig.~\ref{fig:z6linemaps}). However, most of the CGM is completely unaffected. We can see many cold, dense filaments delivering gas to the centre of the halo, fuelling the galaxy's continuing high star formation rate. This is the feeding mechanism for supermassive black holes suggested by a number of other works \citep[e.g.][]{DiMatteo2012,Dubois2012b,Feng2014,Smidt2018}. Because of the likely overproduction of stars in these runs, we still see strongly outflowing supernovae-driven winds expelling high-metallicity gas from the central galaxy into the CGM. 

In the Reference run (right-hand column), we see a significant difference. Some of the dense, inflowing filamentary structures have been disrupted, as shown in the top panel. The significant difference in densities with \fable\ however lies in the more diffuse component, which is much more widespread due to its ejection from the centre of the halo, and is at a higher temperature. This has the additional effect of ejecting cool, dense clumps as well, seen most clearly at angles away from the cosmic filament the halo is embedded in. This has interesting consequences for observations of such gas in H\textsc{i} and Ly$\alpha$, discussed later. 

The earlier growth of the black hole in the Reference run, and subsequent feedback episodes, drive strong, hot outflows (seen in the middle rows), of up to 1000\,km\,s$^{-1}$, that are much faster than in the \fable\ run. This pushes metals out to well beyond $R_\mathrm{vir}$, in a less spherical manner than in the NoAGN/\fable\ runs. The overall metallicity of the CGM is higher in the Reference run, despite a smaller stellar mass of the central galaxy, where the majority of metals are produced. The amount of metals (and detailed metallicity composition) in the high-redshift CGM could therefore potentially be a valuable diagnostic of the growth history of the halo and its black hole. The detection of large patches of metal-enriched gas (with $\gtrsim 0.1\, \mathrm{Z_\odot}$) found at significant distances from high-redshift quasars would lend support to the idea of early enrichment and feedback by such a ``gargantuan'' black hole.

We now look in more detail at the observable properties of the gaseous halo, starting with H\textsc{i} column densities in the top row of Fig.~\ref{fig:z6linemaps}. In these maps, the impact that the \fable\ run begins to have by $z=6$ is more obvious, as it has a broader area covered by high H$\textsc{i}$ column densities. The white contours in these panels show the boundary of Damped Ly$\alpha$ systems (DLAs, $N_\mathrm{H\textsc{i}} = 10^{20.3}$\,cm$^{-2}$). We see therefore how the feedback in the \fable\ run broadens the extent of the central DLA. In the Reference run the central DLA is also broader than in the NoAGN run, though we note that the H\textsc{i} column density distribution is much clumpier in this run. Outside of the main DLA, we see there are a significant number of smaller blobs of gas at DLA column densities which are not present in the NoAGN and \fable\ runs. The contours also show how inflowing filaments, which are clear, coherent structures in the NoAGN and \fable\ runs, become broader and more disturbed. 

In the bottom row of Fig.~\ref{fig:z6linemaps} we show maps of collisional Ly$\alpha$ surface brightness, calculated using the method described in Section \ref{Section:LyaMethods}. As a reminder, full radiative transfer and photoionisation calculations are beyond the scope of this paper and can be investigated in future work, but in this work we can still investigate the impact of different AGN models on the Ly$\alpha$ distribution while making the simplistic assumption of collisional excitation. We note, therefore, that a full treatment of recombination and dust absorption would affect the absolute brightness of Ly$\alpha$ haloes, which should be treated with caution. However, we can probe the relative changes between runs, which can tell us how feedback can affect such observables. 

The total luminosity of the Ly$\alpha$ halo changes between the different feedback runs; the total luminosities in the maps of Fig.~\ref{fig:z6linemaps} are $7.64\times10^{43}$, $7.90\times10^{43}$, and $1.19\times10^{44}$\,erg\,s$^{-1}$ for the NoAGN, \fable, and Reference runs, respectively. Thus the impact of strong, early AGN feedback is to increase the luminosity of the Ly$\alpha$ halo by $\sim\!50$ per cent over a run without any AGN feedback. Looking at the maps, the NoAGN and \fable\ runs again appear largely similar, with the recent feedback in \fable\ not sufficient enough to change the collisional Ly$\alpha$ emission outside of the very centre of the halo. In the Reference run however there is a significant difference, with much brighter Ly$\alpha$ emission in the centre of the halo. Further out in the CGM, we see additional clumps that are bright in Ly$\alpha$, corresponding to the regions of high H\textsc{i} column density. In the radial velocity maps in the central row of Fig.~\ref{fig:z6maps}, we can see that these clumps are visible and are outflowing, albeit at a lower velocity than the surrounding low-density gas. This highlights the importance of strong, and early, black hole feedback in distributing multiphase gas in the CGM of these quasars. Detection of separate Ly$\alpha$ clumps outside the central galaxy could therefore be a signature of such violent and early ejective feedback. 

\subsubsection{The SZ effect}

To further investigate the ejective nature of feedback from ultramassive black holes, and make predictions for upcoming observational experiments, we turn to maps of the SZ effect in Fig.~\ref{fig:z6SZmaps}. In the top row of panels we show the Compton-$y$ parameter of the thermal SZ effect, and we note how in NoAGN run (left), the signal is very bright and centrally concentrated. In the \fable\ run (centre), we can start to see how AGN feedback ejects gas from the very centre of the halo, leading to a reduction in signal strength. Outside of the central 10s of kpc however, the signal is largely unchanged. In contrast, the Reference run shows significant differences, out to and even beyond the virial radius. The central signal is much lower in this run, as the powerful, early feedback ejects significant amounts of gas out into the CGM and this effect wins over the halo gas heating by the quasar feedback. The signatures of past feedback episodes are visible as discontinuities in the signal - shocks - that would be telltale signs of redistributive AGN feedback if they could be detected (though this would require very high angular resolution). 

In the middle row we show the strength of the kinetic SZ signal, which in principle is a powerful probe of the kinematics of the hot halo. Again, there is little obvious difference between the NoAGN and \fable\ runs, with only a slight change in the kinematics in the centre of the halo due to recent feedback episodes. Differences in the Reference run are harder to see visually than in the thermal SZ maps, though we note that the extent of the hot halo has increased with earlier feedback driving strong outflows. The strength of the signal is also somewhat stronger inside the halo due to faster outflows, as shown in Fig.~\ref{fig:z6maps}. 

In the bottom panel, we show the combined SZ signal in the form of a decrement - the difference in energy on top of the near-uniform cosmic microwave background. Note that we have chosen to show the decrement as positive. These maps have also been smoothed with a Gaussian filter on a scale of 1 arcsecond, to mimic the kind of angular resolution that is achievable with modern interferometers like ALMA. The three runs are significantly different in the centre of the haloes. The strongest signal would come from a model without AGN feedback at all (left), which peaks at a strength of $\sim\!1\mu$\,Jy\,arcsec$^{-2}$. The feedback in \fable\ starts to have an impact, lowering the central pressure of the halo by expelling gas, and so leading to a weaker signal. In the Reference run this trend continues and becomes more dramatic, with a significantly lower signal strength in the halo centre. The signal is stronger at larger radii in the Reference run due to the redistribution of gas. SZ observations of high-redshift quasars, and especially those hosting ultramassive black holes like J0100+2802, could therefore be a \textit{key} test of the presence and power of ejective black hole feedback in the early Universe and should be pursued, both with current instruments like ALMA \cite[for recent efforts see e.g.][]{Brownson2019,Jones2023}, and future SZ surveys, such as CMB-HD.

\subsubsection{X-ray emission} \label{Section:XRayResults}

\begin{figure*}
    \centering
    \includegraphics[width=\linewidth]{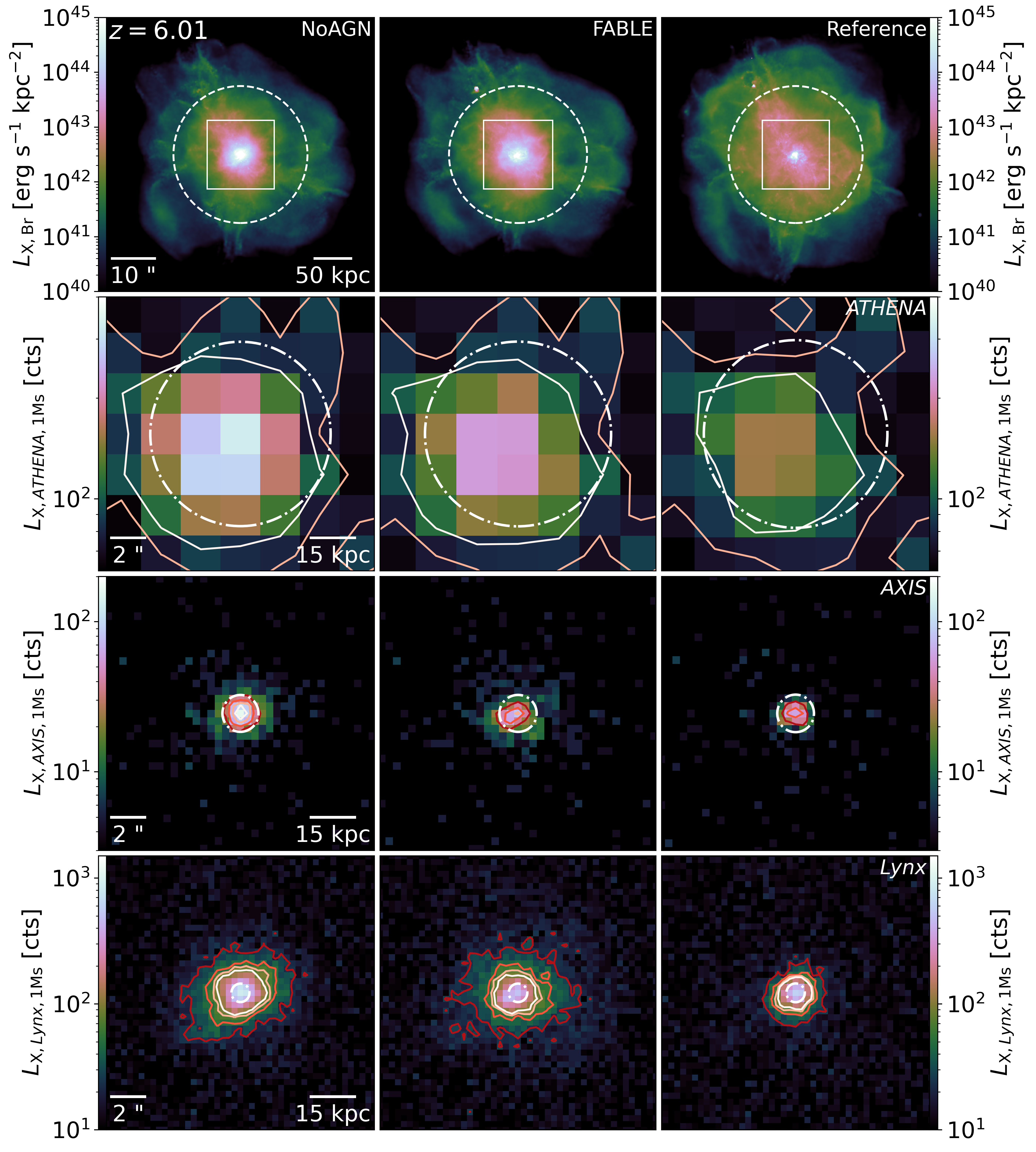}
    \caption[X-ray maps at $z=6$ of the gaseous halo hosting the largest supermassive black hole in our simulations.]{X-ray maps at $z=6$ showing, from top to bottom, X-ray emission using a simple Bremsstrahlung approximation, and mock 1\,Ms exposures with \textit{ATHENA}, \textit{AXIS} and \textit{Lynx}. From left to right the columns show the NoAGN, \fable, and Reference runs. The dashed circle in the top panel shows the virial radius, and the square shows the region that we show mock observations from in the remaining rows. White dot-dashed circles in the centre of the images in the bottom three rows represent the planned point spread function of each X-ray telescope. Contours show the regions from which we detect 25, 50, 75 and 100 counts, going from dark to light. The NoAGN and \fable\ runs have a bright X-ray emitting gaseous halo, which is shrunk significantly in the Reference run due to lower central densities. Only \textit{Lynx} would be able to detect the extended gaseous halo in the Reference run, though any X-ray measurements could potentially place constraints on the impact of AGN feedback.}
    \label{fig:z6Xraymaps}
\end{figure*}

A similar story emerges when looking at mock X-ray observations, which we show in Fig.~\ref{fig:z6Xraymaps}. We note that all of the images in this figure assume a perfect removal of the X-ray emission from the quasar itself. The central, bright spot in the centre of many of these images is instead hot, dense gas, heated by a combination of gravitational heating and feedback (though as a reminder, we remove the central kpc of gas when making mock observations, as this is particularly sensitive to recent bursts of feedback, see Section~\ref{Section:XrayMethods}). For indicative purposes, in the bottom three rows we show a smaller white circle in the centre of each image which indicates the expected point spread function (PSF) of each instrument. 

In the top row, we show maps at the simulation resolution where we make the Bremsstrahlung approximation (see Section~\ref{Section:XrayMethods}). We can see a progression in central X-ray brightness, from the NoAGN run on the left, through the \fable\ run which has some redistribution of gas, to the Reference run which has a significantly different surface brightness profile. The ejected gas in the Reference run significantly raises the X-ray luminosity at large radii, out to and beyond the virial radius, at the expense of the reduced brightness at intermediate radii.  

In the second, third and fourth rows of Fig.~\ref{fig:z6Xraymaps} we show mock 1\,Ms observations (for the method, see Section~\ref{Section:XrayMethods}), using the upcoming and proposed telescopes \textit{ATHENA}, \textit{AXIS} and \textit{Lynx}, respectively. We note that a full analysis of this data would involve binning the data into a surface brightness profile, however for simplicity we just show the mock images. 

With \textit{ATHENA} (second row), the centre of the halo of this galaxy would be bright enough to be detected in all three runs, however the coarse angular resolution of \textit{ATHENA} (expected PSF of 5" at time of writing) means it would be very difficult to disentangle such emission from the light of the quasar itself. With the proposed \textit{AXIS} probe (third row), extended emission could potentially be detectable in the NoAGN and \fable\ runs with a 1\,Ms exposure, though it would be hard to distinguish between them. The lower central densities of the Reference run would make it extremely difficult to pick out extended emission from the background noise with data from \textit{AXIS}. 

The only proposed telescope that could confidently detect the X-ray emission from the gaseous halo itself, even in the Reference run, is \textit{Lynx}. In the bottom row we can see how the NoAGN and \fable\ runs would themselves have a slightly different morphology, with the potential to distinguish between them and detect further extended emission with a surface brightness profile. While the total X-ray brightness in the extended hot halo of the Reference run is significantly lower due to reduced central densities, a long exposure from \textit{Lynx} could still detect this and distinguish it from the PSF of the quasar itself. 

Despite only \textit{Lynx} being able to detect the gaseous halo in the Reference run, observations from any of the upcoming or proposed X-ray telescopes could provide interesting constraints on the amount of ejective feedback that could have taken place in high-redshift quasars. And while spatially resolving quasar host haloes is beyond the capabilities of other imminent and proposed future X-ray probes like \textit{XRISM} and \textit{LEM}, their high spectral resolution could also allow a detailed study of the quasar surroundings using the light of the quasar itself.

\section{Discussion} \label{Section:Discussion}

\subsection{Caveats of our modelling}

There are a number of caveats to work investigating high-redshift quasars, both in our work and across other simulations, that are worth discussing. 

Firstly, the seeding of initial black holes in our models (and most cosmological simulations) is very simplistic. The $10^5 \, h^{-1}$\,M$_\odot$ seed mass used in this work is on the larger end of theoretical seed masses, and would originate from a direct collapse black hole (DCBH). With the changes made to the Reference run, a $10^5 \, h^{-1}$\,M$_\odot$ black hole is seeded at $z\sim18$. DCBHs of $10^6$\,M$_\odot$ are potentially possible, which could go some way to explaining the early growth of high-redshift quasars. Smaller, stellar mass seeds that form sufficiently early might be able grow to become $\sim\! 10^5$\,M$_\odot$ by the seed time of the Reference run ($z\sim18$), however these may suffer from low initial growth due to being in small haloes \citep[see e.g.][]{Inayoshi2020}. Clues to the early formation of black holes that go on to become high-redshift quasars will be unravelled with the upcoming gravitational wave observatory \textit{LISA}, which will be able to detect mergers between intermediate mass black holes out to extremely high-redshift. We will therefore be able to investigate how important black holes mergers are in this early growth phase, which will then allow us to constrain the seeds of high-redshift quasars. Before \textit{LISA} becomes operational, crucial clues into the demographics of black holes seeds may come from \textit{JWST} observations.

The second caveat to modelling black hole growth in cosmological simulations is that of accretion. We, like many other simulations, use a Bondi-based accretion model. This model assumes spherically symmetric accretion onto a compact object, which may not be the case for many (most) black holes (especially at high redshift where galactic accretion tends to be preferentially along filaments, though this ultimately depends sensitively on the gas morphology on small scale close to the accretion disc). To try and account for the unresolved nature of black hole accretion, a boost factor $\alpha$ is commonly employed to artificially increase the accretion rate ($\alpha=100$ in Illustris and \fable). However there is considerable uncertainty in the validity of the Bondi model, and alternative models have been proposed that could change the early growth of black holes \citep[such as torque-based accretion or alpha disc models, e.g.][]{HopkinsQuataert2011,Fiacconi2018,Dave2019,AnglesAlcazar2021,Talbot2021}. A further caveat to the accretion model in this work comes from limiting the accretion rate to (a factor of) the Eddington rate. In the Reference run we relax the assumption of Eddington-limited accretion, by a factor of two, and find it helps to increase the final black hole mass. This agrees with the findings of \citet{Zhu2022}, who further investigate Eddington limits of $5$ and $10^4$ and find the opposite effect due to increased feedback energy \citep[though black holes can still effectively grow if feedback is further weakened, see e.g.][]{Massonneau2023}. In reality, tighter observational constraints on the accretion rates and masses of black holes will be needed to test our models of accretion at high redshift, and what we present in this work is just one potential pathway to growing a `gargantuan' black hole in the early Universe. 

\section{Conclusions} \label{Section:Conclusions}

Hundreds of quasars have now been detected at $z \gtrsim 6$, with many of them even at $z \gtrsim 7$ \citep[e.g.][]{Mortlock2011,Banados2018,Yang2020,Wang2021}. Among these, the record holder in terms of the inferred black hole mass is J0100+2802, at a hefty $1.24\times10^{10}\,\mathrm{M}_\odot$
at $z=6.30$ \citep{Wu2015}. 

To study the impact of such `gargantuan' black holes on their surroundings, we have performed three zoom-in simulations of the largest halo in the $500\,h^{-1}$\,Mpc Millennium simulation box \citep{Millennium} to investigate the early growth of supermassive black holes and, importantly, their impact on their surroundings. In doing so, we aimed to reproduce black holes with masses approaching the largest known high-redshift quasars, as this extreme mass regime has been largely unexplored with cosmological simulations of structure formation \citep[for one recent attempt, see][]{Zhu2022}. This stems from the difficulty in growing such massive black holes in less than a billion years of cosmic time even when starting from very massive seeds \citep{Sijacki2009,DiMatteo2012,Costa2014}. We expect, however, based on their extreme luminosities of $\sim 10^{48}\, \mathrm{erg}\,\mathrm{s}^{-1}$, that these objects may have a profound impact on their host galaxy and halo.

We have shown how the fiducial \fable\ model is unable to generate such a black hole early enough in the simulation. The \fable\ model is also insufficient to regulate star formation at high redshift, and furthermore, many properties of the host galaxy and CGM are indistinguishable between the \fable\ model and a run without AGN at all. This is the case even though at low redshift, \fable\ produces reasonable stellar masses and star formation rates in massive galaxy clusters \citep{FABLE1}. The properties of high-redshift quasar hosts could therefore provide further important constraints on theoretical models of AGN feedback and their redshift evolution. 

With changes we made to the base \fable\ simulation, primarily through seeding black holes earlier, allowing mildly super-Eddington accretion, and reducing AGN feedback efficiency (but note that the total injected AGN feedback power depends crucially on black hole mass as well), we have grown a supermassive black hole with a mass in excess of $10^{10}\,\mathrm{M}_\odot$ by $z=6$, in agreement with the largest estimated black hole mass from observations of bright quasars at that redshift \citep{Wu2015}. This is the only simulation from the six most massive Millennium halos at $z = 6$ in which such an ultramassive black hole forms, highlighting the high rarity of these objects and unique conditions that need to be met, such as having a very early assembly history. The bolometric luminosity of this black hole often sits at $10^{47-48}\,\mathrm{erg\,s}^{-1}$, matching the inferred bolometric luminosities of such observed quasars. 

Despite sustained, high luminosities, the quasars in both the \fable\ and Reference runs spend much of their life with a significant number of sightlines around them obscured (at least in the UV) due to a large amount of surrounding gas and dust. In the Reference run, more powerful quasar feedback is able to more effectively clear some of this dense gas earlier and more effectively, leading to a higher probability of it being observed. It is worth noting that even with strong quasar feedback there are a number of sighlines with large infrared optical depths, highlighting that the radiation pressure-driven outflows are likely an important feedback mechanism of such highly obscured quasars. Significant amounts of obscuration also seem required to explain the very short inferred quasar lifetimes from observations of proximity zones. We find that without efficient feedback from the quasar expelling gas from the halo centre, a central quasar would remain highly obscured in the rest-frame UV throughout. 

We have shown how such massive black holes drive powerful outflows in high-redshift galaxies, significantly lowering central stellar and gas densities and pressures. These outflows can eject metals to large distances from the central galaxy at $z>6$, which could be a way to probe and constrain such models with future observations. We also find that early quasar feedback can significantly enhance the Ly$\alpha$ luminosity of its host halo. The key factor in all of this seems to be how early a black hole can grow massive enough relative to the potential well depth of its host halo. If an AGN can start to drive outflows when the potential well of the dark matter halo is shallower, the feedback will be more effective. 

Tests for such early feedback and enrichment models will come from a number of angles in the coming decade. At optical and infrared wavelengths, the Vera Rubin Observatory, \textit{Euclid}, and the \textit{Roman Space Telescope} will be capable of discovering a larger sample of high-redshift quasars, as SDSS has done in the previous decade. The ongoing science missions of ALMA and \textit{JWST} will also allow the study of high-redshift galaxies, with insights into their stellar and gaseous components and the detection of high-velocity outflows. We have shown how the star formation rate and estimated dust mass of our massive quasar in the Reference run are in good agreement with those inferred from observations of J0100+2802. Moreover, \textit{JWST} is already starting to probe fainter AGN at very high redshifts which will give us valuable constraints on the growth of progenitors of these ultramassive $z = 6$ quasars, and even on black hole seeding models. 

We have shown how the SZ effect can be used to investigate the impact of the growth of high-redshift ultramassive black holes on their surroundings, meaning future instruments, like the Simons Observatory, CMB-S4, CMB-HD, and even state-of-the-art current observations with ALMA, could provide further pieces of this puzzle.

In X-rays, we find that observations from the future \textit{Athena} mission, or the proposed probe \textit{AXIS}, would be able to place constraints on the impact black hole feedback can have on the CGM of high-redshift quasars, though still detecting the hot halo in our most extreme case of feedback would be very difficult. Doing so would require the capabilities of the proposed \textit{Lynx} observatory, whose approval and launch would allow the study of high-redshift quasars and their hosts in exquisite detail. 

The exact mechanisms and impacts of black hole growth in the early Universe are still very uncertain, but with the wealth of new observational data becoming available now and in the near future, progress on these fundamental questions should be possible. Simulations like those presented in this paper, that make predictions for a range of observables that can be directly correlated with supermassive black hole masses and luminosities at high redshift, are therefore timely. Despite some of the dramatic changes in results we find between our runs, the modifications we made to the \fable\ model are relatively moderate in nature, especially considering the large uncertainties in black hole seeding and accretion, which bodes well for constructing next-generation models that will be able to produce these `gargantuan' black holes while maintaining the lower-redshift successes of existing models. 

It is worth emphasising here that our modified \fable\ setup chiefly promotes earlier black hole growth and, once AGN feedback becomes effective, the differences between the models are likely to get reduced over cosmic time due to effective self-regulation. Interestingly, with such `gargantuan' black holes we may reach the regime where further radiatively efficient growth may not be possible \citep[see e.g.][]{King2016}. This could lead to ultramassive but inactive black holes at lower redshift \citep[which could potentially be detected through gravitational lensing, e.g.][]{Nightingale2023}, which would be timely to explore with detailed numerical simulations. Hence, with further theoretical developments of these models, and a better understanding of the early evolution of gaseous haloes around galaxies, these predictions can be refined and compared to new data to gain crucial insight into the emergence of these extreme and fascinating compact objects in the early Universe.   

\section*{Acknowledgements}
The authors would like to thank Colin DeGraf, Martin Haehnelt, Roberto Maiolino, Yueying Ni, Chris Reynolds, Jan Scholtz, Aaron Smith, and John ZuHone for helpful conversations and comments about this work.
This work was supported by the Simons Collaboration on “Learning the Universe”. JB and CW acknowledge support from the Science and Technologies Facilities Council (STFC) for PhD studentships funded by UKRI grants ST/S5053-4/1 and 2602262, respectively. JB and DS acknowledge support from ERC Starting Grant 638707 `Black holes and their host galaxies: co-evolution across cosmic time'. 
This work was performed using resources provided by: the DiRAC@Durham facility managed by the Institute for Computational Cosmology on behalf of the STFC DiRAC HPC Facility (www.dirac.ac.uk). The equipment was funded by BEIS capital funding via STFC capital grants ST/P002293/1, ST/R002371/1 and ST/S002502/1, Durham University and STFC operations grant ST/R000832/1. DiRAC is part of the National e-Infrastructure; the Cambridge Service for Data Driven Discovery (CSD3) operated by the University of Cambridge Research Computing Service (www.csd3.cam.ac.uk), provided by Dell EMC and Intel using Tier-2 funding from the Engineering and Physical Sciences Research Council (capital grant EP/P020259/1). This work made use of the NumPy \citep{Numpy}, SciPy \citep{SciPy}, and Matplotlib \citep{Matplotlib} Python packages.

%%%%%%%%%%%%%%%%%%%%%%%%%%%%%%%%%%%%%%%%%%%%%%%%%%
\section*{Data Availability}
The data used in this work may be shared on reasonable request to the authors.



\bibliographystyle{mnras}
\bibliography{references} % if your bibtex file is called example.bib

%\begin{thebibliography}{99}
%\end{thebibliography}

%%%%%%%%%%%%%%%%%%%%%%%%%%%%%%%%%%%%%%%%%%%%%%%%%%

%%%%%%%%%%%%%%%%% APPENDICES %%%%%%%%%%%%%%%%%%%%%

\appendix

\section{Black hole growth in other massive Millennium haloes} \label{AppA}

\begin{figure}
    \centering
    \includegraphics[width=\linewidth]{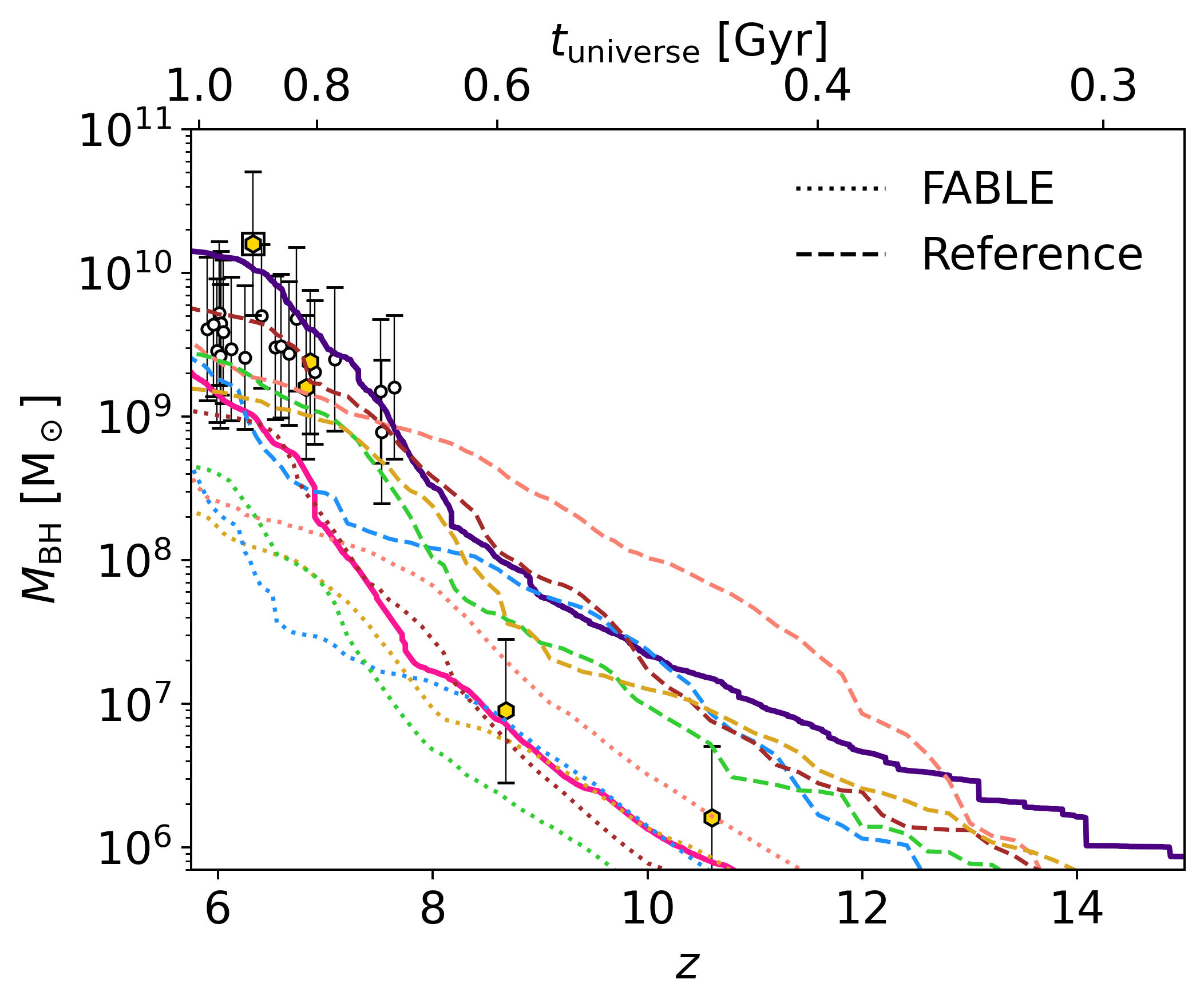}
    \caption[Black hole mass growth curve in different haloes.]{Growth of black hole mass as a function of redshift for the largest black hole in a six different haloes. Individual points show where observations of black holes lie, and are the same as in Fig.~\ref{fig:BHgrowth}. Error bars again show a representative $0.5$\,dex error. The black square highlights the data from J0100+2802, the largest observed high-redshift quasar. The \fable\ and Reference runs from Fig.~\ref{fig:BHgrowth} are reproduced in the solid lines. Dotted and dashed lines show \fable\ and Reference runs for five other haloes.}
    \label{fig:AppendixBHGrowth}
\end{figure}

To explore the rarity of our most massive black holes in the \fable\ and Reference setups, we simulate additional five Millennium haloes, which together with the simulated system presented in the main body of the paper, comprise the six most massive Millennium haloes at $z = 6$. We take the initial conditions identical to the ones employed by \citet{Costa2014} and adopt the same simulation setup and physics implementation as in our Section~\ref{Section:FABLE} and \ref{Section:Modifications}, for the \fable\ and Reference setup, respectively. The virial halo masses at $z = 6$ span from $2.9 \times 10^{12}\,\mathrm{M}_\odot$ to $5.6 \times 10^{12}\,\mathrm{M}_\odot$ \citep[for further details see table 2 of][runs labelled O1 to O6]{Costa2014}. Figure~\ref{fig:AppendixBHGrowth} shows the growth of the most massive black holes in these six Millennium haloes for the \fable\ setup (dotted) and the Reference setup (dashed). The \fable\ and Reference runs from Fig.~\ref{fig:BHgrowth} are reproduced (solid) together with the observational data. On average black hole masses in the \fable\ setup are one order of magnitude lower than those in the Reference setup, which match observational data. Only in the most massive halo of the Millennium simulation at $z = 6$, which has an exceptionally early assembly history does a `gargantuan' black hole with $\sim 10^{10}\,\mathrm{M}_\odot$ form. 

We note that several haloes with the Reference model produce black holes with masses comparable to even higher redshift quasars ($z>7.5$), though these black holes are then less massive by $z=6$ than the main halo used in this work. This indicates how the particular growth history of a halo can have a significant impact on how large a black hole the halo can host at a given time.

%%%%%%%%%%%%%%%%%%%%%%%%%%%%%%%%%%%%%%%%%%%%%%%%%%

% Don't change these lines
\bsp	% typesetting comment
\label{lastpage}
\end{document}